\documentclass{aa}
\usepackage{physics}
\usepackage{txfonts}
\usepackage{graphicx}
\usepackage{placeins}
\graphicspath{{figures/}}
\usepackage[separate-uncertainty=true,multi-part-units=single]{siunitx}
\usepackage{wasysym}
\DeclareSIUnit{\mass}{M}
\DeclareSIUnit{\radius}{R}
\DeclareSIQualifier{\solar}{\astrosun}
\DeclareSIQualifier{\stellar}{_*}
\usepackage{amsmath}
\DeclareMathOperator{\Cov}{Cov}

\newcommand\orc[1]{\href{https://orcid.org/#1}{\includegraphics[width=3mm]{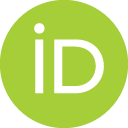}}}

\title{On the feasibility of structure inversions for gravity-mode pulsators}
\author{
  V. Vanlaer\inst{\ref{kul}}\orc{0000-0003-4923-6199},
  C. Aerts\inst{\ref{kul},\ref{imapp},\ref{mpiaheidelberg},\ref{CCA}}\orc{0000-0003-1822-7126},
  E. P. Bellinger\inst{\ref{mpagarching}}\orc{0000-0003-4456-4863},
  J. Christensen-Dalsgaard\inst{\ref{aarhus}}\orc{0000-0001-5137-0966}
}
\abstract{Gravity-mode asteroseismology has significantly improved our understanding of mixing in intermediate mass stars.
However, theoretical pulsation periods of stellar models remain in tension with observations, and it is often unclear how the models of these stars should be further improved.
Inversions provide a path forward by directly probing the internal structure of these stars from their pulsation periods, quantifying which parts of the model are in need of improvement.
This method has been used with success in the case of solar-like pulsators, but has not yet been applied to main-sequence gravity-mode pulsators.}
{Our aim is to determine whether structure inversions for gravity-mode pulsators are feasible.
We focus on the case of slowly rotating slowly pulsating B-type (SPB) stars.}
{We computed and analyzed dipole mode kernels for three variables pairs: $(\rho, c)$, $(N^2, c)$, and $(N^2, \rho)$.
We assessed the potential of these kernels by predicting the oscillation frequencies of a model after perturbing its structure. We then tested two inversion methods, regularized least squares (RLS) and subtractive optimally localized averages (SOLA), using a model grid computed with the MESA stellar evolution code and the GYRE pulsation code.}
{We find that changing the stellar structure affects the oscillation frequencies in a nonlinear way.
The oscillation modes for which this nonlinear dependency is the strongest are in resonance with the near-core peak in the buoyancy frequency.
The near-core region of the star can be probed with SOLA, while RLS requires fine tuning to obtain accurate results.
Both RLS and SOLA are strongly affected by the nonlinear dependencies on the structure differences, as these methods are based on a first-order approximation.
These inversion methods need to be modified for meaningful applications of inversions to SPB stars.}
{Our results show that inversions of gravity-mode pulsators are possible in principle, but that the typical inversion methods developed for solar-like oscillators are not applicable. Future work should focus on developing nonlinear inversion methods.}

\date{Received 1 December 2022 / Accepted 2 May 2023}
\institute{Institute of Astronomy, KU Leuven, Celestijnenlaan 200D, 3001, Leuven, Belgium \label{kul}\\
  \email{vincent.vanlaer@kuleuven.be} \and
  Department of Astrophysics, IMAPP, Radboud University Nijmegen, PO Box 9010, 6500 GL Nijmegen, The Netherlands \label{imapp} \and
  Max Planck Institute for Astronomy, K\"onigstuhl 17, 69117, Heidelberg, Germany \label{mpiaheidelberg} \and
  Guest Researcher, Center for Computational Astrophysics, Flatiron Institute, 162 Fifth Ave, New York, NY 10010, USA \label{CCA} \and
  Max Planck Institute for Astrophysics, Karl-Schwarzschild-Str. 1, 85748 Garching, Germany \label{mpagarching} \and
  Stellar Astrophysics Centre, Department of Physics and Astronomy, Aarhus University, Denmark \label{aarhus}
}
\authorrunning{Vanlaer et al.}
\bibpunct{(}{)}{;}{a}{}{,}

\begin{document}
\maketitle

\section{Introduction}

The main-sequence evolution of intermediate-mass stars (1.3-8\SI{}{\mass\solar}) is strongly dependent on the size and mass of the central convection zone.
However, the properties of the boundary between this convection zone and the radiative envelope are largely unknown, affecting predictions of the age and final core mass of the star \citep[e.g.,][for summaries]{johnston-one-2021,pedersen-internal-2021,pedersen-diversity-2022}.
The main challenge is determining the degree of element mixing that occurs in these boundary layers and throughout the envelope of the star, as well as how this mixing changes as the star evolves.
Different processes contribute to this mixing: overshooting near the convection zone, differential rotation, settling and levitation of chemical species, and internal waves.
Asteroseismic observations have made the properties of these deep interior layers accessible (see \citealt{aerts-probing-2021} for a review).
In order to derive these properties, one typically fits predicted observables from a grid of stellar models computed with a stellar evolution code, to observations of the star, including asteroseismic data, but also additional constraints from spectroscopy or astrometry. 
We refer to this way of modeling as forward modeling.

While forward modeling has been applied with success to tens of $\gamma$ Dor stars \citep{mombarg-asteroseismic-2019,mombarg-constraining-2021} and slowly pulsating B-type stars \citep[SPB stars,][]{degroote-deviations-2010,pedersen-internal-2021,szewczuk-seismic-2022}, this procedure requires computationally expensive model grids \citep{aerts-forward-2018}.
Furthermore, it may be the case that certain physical processes are not accurately modeled or not even considered at all, even though they are necessary to accurately describe the evolution of a star.
This not only leads to many parameters that can be tuned, but also to differences in the implementation of physical phenomena between stellar evolution codes such as MESA \citep{jermyn-modules-2023}, the Geneva Code \citep{eggenberger-geneva-2008}, GARSTEC \citep{weiss-garstec-2008}, FRANEC \citep{deglinnocenti-franec-2008}, Monash \citep{lattanzio-asymptotic-1986}, and STAREVOL \citep{forestini-new-1991} among others.

A method that can help alleviate these challenges is a (structure) inversion of the star we want to model \citep[e.g.,][]{gough-inversion-1991,bellinger-inverse-2020}.
An inversion attempts to reconstruct the internal structure of the star directly from observed and identified oscillation modes.
Since the oscillation frequencies on their own do not provide complete information about the internal structure of the star, it is necessary to use a reference model as a starting point for the inversions.
The differences between the oscillation frequencies of the reference model and the observations can then be used to infer inaccuracies in the structure of the reference model.
It is in principle possible to quantify these inaccuracies as long as the differences between the stellar model and the actual star are small (such that the differences in frequency are linearly related to the differences in structure) and sufficient oscillation modes are observed.
This method has been applied successfully with high accuracy to the sound speed and rotation profile of the Sun \citep[e.g.,][for a review]{basu-global-2016,buldgen-global-2019, christensen-dalsgaard-solar-2021}.

For stars beyond our own Sun, this kind of analysis became possible with the advent of space telescopes such as \textit{Kepler} \citep{borucki-kepler-2010,koch-kepler-2010} and the Transiting Exoplanet Survey Satellite \citep[TESS,][]{ricker-transiting-2014}.
Thus far, inversions based on solar-like oscillations (p modes; sometimes mixed modes) have been applied to the sound speed profiles in the central regions of main-sequence stars \citep{bellinger-modelindependent-2017,bellinger-testing-2019,buldgen-thorough-2022}, rotation profiles of main-sequence stars, subgiants and red giants \citep{deheuvels-seismic-2012,deheuvels-seismic-2014,dimauro-internal-2016,dimauro-rotational-2018,triana-internal-2017,hatta-twodimensional-2019,hatta-bayesian-2022}, including exoplanet hosts \citep{buldgen-coralie-2022,betrisey-kepler93-2022}, the central structure of a subgiant star \citep{kosovichev-resolving-2019, bellinger-asteroseismic-2021}, and various global parameters of these stars, such as the mean density \citep{reese-estimating-2012,buldgen-constraints-2016,buldgen-indepth-2016,buldgen-mean-2019}.

Inversions of g-mode pulsators have only been applied to the rotation profile of two pre-white dwarfs \citep{kawaler-prospects-1999,corsico-probing-2011} and two main-sequence stars, a $\gamma$ Dor, $\delta$ Sct hybrid pulsator \citep{kurtz-asteroseismic-2014} and an SPB star \citep{triana-internal-2015}.
Some early feasibility studies on white dwarfs \citep{shibahashi-white-1988,takata-seismic-2001} have attempted g-mode inversions.
However, their results showed that given the limited amount of oscillation modes available at the time, inversions would not be possible.
Structure inversions for g modes have not yet been attempted since then, in part due to g-mode pulsators having their own unique challenges compared to p-mode pulsators.
In general, a larger number of oscillation frequencies can be identified in p-mode pulsators: on average $\sim$45 modes for solar-like pulsators \citep{lund-standing-2017}, while $\sim$20 for SPB stars \citep{pedersen-internal-2021}, and $\sim$30 for $\gamma$ Dor pulsators \citep{vanreeth-gravitymode-2015}.
The g modes are determined predominantly by the buoyancy frequency, $N(r)$, which is strongly peaked and therefore more difficult to invert for compared to inversions for the sound speed or the density.
However, the buoyancy frequency is the main property of the star linking the g modes to the temperature and composition gradients in the star \citep{aerts-asteroseismology-2010,michielsen-probing-2021}, making it the main target of our inversions.
The proximity of the Sun has also allowed us to constrain its internal structure much better than that of other stars, making it an excellent starting point for other solar-like stars \citep{christensen-dalsgaard-helioseismology-2002,christensen-dalsgaard-solar-2021}.
Such a starting point is not available for g-mode pulsators with a fully developed convective core.
For example, while the mass, radius, and luminosity of the Sun have uncertainties less than 0.1\% (they are essentially known perfectly for the purpose of stellar evolution), the best estimates for main-sequence g-mode pulsators have uncertainties of 5 to 10\% \citep{pedersen-diversity-2022,mombarg-constraining-2021}.
Finally, many of the g-mode pulsators with a convective core have frequencies within the gravito-inertial regime, where rotation cannot be treated perturbatively \citep{aerts-rossby-2021} and zonal modes are not the norm \citep{pedersen-internal-2021}.
Hence, effects of rotation need to be included in a complete inversion method for g-mode pulsators.

In this paper, we explore the first steps that need to be taken to develop a full inversion method for g-mode pulsators, considering only slowly rotating stars.
We explain how structure inversions work in Sect.~\ref{sec:inversion} and discuss two methods that have been used for solar-like pulsators: regularized least squares (RLS) and subtractive optimally localized averages (SOLA).
In Sect.~\ref{sec:model}, we introduce the stellar models used for testing the inversions.
Section~\ref{sec:avoided-crossings} discusses how avoided crossings limit the applicability of linear inversions.
In Sect.~\ref{sec:test} we analyze which oscillation kernels are most suitable for g-mode inversions and how well the two inversion methods are able to recover the stellar structure.
Finally, we summarize and conclude in Sect.~\ref{sec:conclusion}.

\section{Methods for g-mode inversions of stellar properties}
\label{sec:inversion}
Changes in the structure of a star (e.g. changes in the density, sound speed, etc.) change the oscillation frequencies and other properties of its pulsation modes.
The goal of an inversion is to reconstruct differences between a stellar model and reality from differences between observed frequencies and modeled frequencies for oscillation modes that can be unambiguously identified in terms of their wave numbers. 
If these differences are small enough, this relation can be approximated linearly.
As the model and the actual star may have different radii, we define the differences in structure at the same fractional radius instead of using the same physical radial coordinate.
This means that we describe the differences in a structure property $x$ between the model and the star or between different models as
\begin{equation} \delta x (r) \equiv x_\mathrm{obs} \left(r \frac{R_\mathrm{obs}}{R_\mathrm{ref}}\right) - x_\mathrm{ref} (r)\,,  \end{equation}
where $R$ is the stellar radius, and the subscripts ``ref'' and ``obs'' refer to the reference model and the observed star, respectively.
Additionally, the derivation of the linear relation is based on the variational principle, which requires a constant stellar radius.
This can be overcome by considering the parameters represented by $x$ dimensionless, scaled according to the radius and mass of the star.
For example, the dimensionless density, $\hat\rho$, sound speed, $\hat c$, and frequency, $\sigma$, are defined as
\begin{equation} \hat \rho \equiv \frac{R^3}{M}\,\rho\,,\quad\hat c \equiv \sqrt{\frac{R}{GM}}\,c\,,\quad\sigma^2 \equiv \frac{R^3}{GM}\,\omega^2\,.  \end{equation}
Since we are comparing dimensionless quantities, the properties of the actual star can only truly be recovered if the radius and mass of the star can also be recovered in the inversion.
In practice, as the oscillation frequencies have a dimension of [time]$^{-1}$, only the ratio $R^3/M$ can be recovered.

For one oscillation mode, indexed by $i$, with angular frequency $\omega_i$, the linear relation is expressed with the following integral equation \citep[e.g.,][]{gough-inversion-1991, basu-stellar-2003}.
\begin{equation}
	\Delta_i \equiv \frac{\delta\omega_i}{\omega_i} \approx \int_0^{R} K_{x_1,x_2,\cdots}^ i \frac{\delta \hat x_1}{\hat x_1}
        + K_{x_2,x_1,\cdots}^i \frac{\delta \hat x_2}{\hat x_2} + \cdots \dd{r}
                                               - \frac{\delta q}{q}\,,
\label{eqn:kernel-definition}\end{equation}
where $x_1, ...$ is some set of independent variables describing the stellar structure.
The parameter $q$ is defined as $\sqrt{R^3/GM}$ and represents the scaling of the oscillation frequencies with the fundamental parameters of the star, making the frequencies dimensionless.
The functions $K_{x_i, ...}^ i$, called the mode kernels, contain all the information on the original model of the star, and depend on the properties of the observed oscillation modes, as indicated by the index $i$. 
Each oscillation mode is influenced differently by the internal stellar properties, hence these kernels will in general be different from each other.
This makes it possible to disentangle the structure differences required to reproduce the observed oscillation frequencies.
To which degree the structure can be disentangled depends on the amount of oscillation modes that can be detected and how independent the oscillation modes are \citep{pijpers-sola-1994}.

Not all choices for the variables $x_1, x_2, ...$ in Eq.~\eqref{eqn:kernel-definition} are meaningful.
Many of the variables describing the internal structure of the star depend on each other via thermodynamic and fluid dynamics equations.
For example, the pressure gradient can be expressed as a function of the density profile using the equation of hydrostatic support.
Just considering the structure of the star (i.e. ignoring the dynamical effects of rotation and magnetism on the oscillations), only two variables are needed to capture all possible changes \citep{gough-inversion-1991}.
An example of such a pair of variables is the density and the sound speed.
The choice of these variables affects the validity of the linear approximations made in Eq.~\eqref{eqn:kernel-definition} \citep{buldgen-analysis-2017}.
Typically, one chooses a pair of variables such that the contribution from one of the variables dominates over the other.
What variable pair is best depends on the type of oscillation modes and other assumptions that are made regarding the structure of the star.
For example, assuming that the equation of state is known, allows one to use the local helium abundance as one of the variables.
In the case of the Sun and other solar-like stars, the oscillation frequencies are insensitive to the local helium abundance except for a small region near the surface, making the kernels for that local helium abundance negligible \citep[e.g.,][]{basu-stellar-2003}.
In the limit of high order, the p-mode oscillation frequency is determined solely by the sound speed, in what is known as Duvall's law, suppressing the relevance of the other variable \citep{duvall-dispersion-1982}.
A similar relation exists for low-degree, high-order g modes in a slowly rotating star, but with the buoyancy frequency $N$ instead of the sound speed \citep{tassoul-asymptotic-1980}.
The buoyancy frequency is also directly linked to the chemical gradient inside the star.
We therefore use the buoyancy frequency as our primary target for g-mode inversions, in combination with either the sound speed or the density.
As a reference, we also use the combination of the density and the sound speed as kernels, as this is a standard kernel pair that forms the starting point for the derivation of other kernel pairs.
Our derivations of the mathematical expressions for the kernel pairs $(N^2, c)$ and $(N^2, \rho)$ can be found in Appendix~\ref{sec:kernel-derivations}.

Equation \eqref{eqn:kernel-definition} allows one to compute the changes in oscillation frequencies from changes in the structure of the star, but the goal of an inversion is to do the opposite.
Starting from differences between modeled and observed oscillation frequencies, we want to determine which changes need to be made to the structure of the star to bring the two into agreement.
However, the structure of the star has infinite degrees of freedom, making a one-to-one relation impossible.
Additionally, we need to determine the parameter $\delta q$, which induces some degeneracies between the structure of the star and the scaling of the star.
In the case of p modes, $\delta q$ can be estimated from their average shift in frequency \citep{roxburgh--1998}, but such a relation does not exist for g modes.
In the next two sections, we describe two methods that can be used to overcome the large number of degrees of freedom of the stellar structure for g-mode inversions.

\subsection{Regularized least squares}

The first method we use is the regularized least squares (RLS) method \citep[e.g.,][]{christensen-dalsgaard-comparison-1990}).
RLS makes two restrictions on the profiles inferred from inversion.
First, the inferred profiles are limited to a set of predefined functions (often constructed from a set of basis functions).
In this case, these functions form a step profile, which can be nonuniform in the radial coordinate to optimally match the resolving power of the oscillation modes at different locations in the star.
Secondly, a term is added to the least squares minimization that penalizes sharp features in the inferred profile.
Given a discretized step profile ${\delta \bar x_1^j}$ (assuming only one variable is considered), we define the predicted frequency differences as
\begin{equation} \bar\Delta_i = \sum_j G_{ij}\,\delta \bar x_1^j\,\text{,\quad where}\quad G_{ij} = \int_{r_j}^{r_{j+1}} \frac{K^i_{x_1, \cdots}}{x_1} \dd{r}\,. \label{eqn:delta-bar} \end{equation}
The RLS method then consists of minimizing
\begin{equation} \mathcal{L} = \sum_i\frac{(\Delta_i - \bar\Delta_i)^2}{\varepsilon^2_i} + \mu \sum_j  h(\delta\bar x_j)\,, \end{equation}
with $\varepsilon_i$ being the uncertainty on the frequency difference $\Delta_i$ and $h$ a function of the inferred profile, defining what we consider sharp features and how much of those sharp features are present in the inferred profile.
This function can be at most a quadratic function of the inferred profile, as otherwise minimizing $\mathcal{L}$ is no longer a linear problem.
While this is not a strict requirement, it allows us to use minimization techniques such as ordinary least squares.
As is often done for the Sun and other solar-like pulsators (e.g. \citealt{christensen-dalsgaard-comparison-1990}), we use the square of the second derivative integrated over the entire profile as the penalty function:
\begin{equation} h(\delta \bar x_1) = \int_0^R \left(\dv{^{2}\delta \bar x_1}{r^{2}}\right)^2\dd{r}\,.\end{equation}
Since these are discrete profiles, we approximate this integral using finite differences.
The regularization parameter $\mu$ controls the importance of the smoothing term, compared to minimizing the residuals.
This factor depends on the number of identified oscillation modes as well as the number of points in the discretized profile.
The optimal value for $\mu$ needs to be determined as part of the fitting procedure.

\subsection{Subtractive optimally localized averages}

Subtractive optimally localized averages (SOLA; \citealt{pijpers-sola-1994}) does not try to fit the data, but rather constructs a kernel, called the averaging kernel, that is localized around a certain region of the star.
This kernel is a linear combination of the kernels of the oscillation modes:
\begin{equation} \mathcal{K} = \sum_i c_i K^i_{x_1, x_2}\,,\end{equation}
with the parameters $c_i$ chosen such that $\mathcal{K}$ is localized around a certain point $r_0$.
What this localization of the averaging kernel means is determined by a target kernel $\mathcal{T}$, which the averaging kernel should approximate as closely as possible.
As is typical for p-mode inversions, we use a modified Gaussian distribution as $\mathcal{T}$ such that it always tends to zero in the center of the star \citep[e.g.,][]{basu-asteroseismic-2017}:
\begin{equation} \mathcal{T}(r; r_0, \Delta) = a\cdot r\exp \left( - \left(\frac{r - r_0}{\Delta} + \frac{\Delta}{2r_0}\right)^2\right)\,, \end{equation}
where $\Delta$ is the width of the Gaussian, $r_0$ the location of the peak of the Gaussian and $a$ a scale factor such that the target kernel integrates to one.
For SOLA the localization is implemented by minimizing
\begin{equation} \mathcal{L} = \int_0^R \left(\mathcal{K} - \mathcal{T}\right)^2\dd{r}\,. \label{eqn:sola-simple}\end{equation}
We also require that the averaging kernel integrates to one, to keep the meaning of the kernel averaging the region around a certain point.
If the averaging kernel is localized around a reference position $r_0$, the relation between the structure of the star and the observations becomes
\begin{equation} \sum_i c_i \Delta_i \approx \int_0^R \mathcal{K} \delta x_1\dd{r} \approx \delta x_1(r_0)\,, \label{eqn:sola-approx} \end{equation}
such that we can use $\sum_i c_i \Delta_i$ as an estimate of $\delta x$ around the position $r_0$.
We note that this is only an estimation of $\delta x_1$ at the point $r_0$ if the actual structure differences are sufficiently smooth.

Ensuring that the averaging kernel is as close as possible to the target kernel is not the only property that the coefficients $c_i$ should have.
First, we need to consider the effect of the second variable in the variable pair ($x_1$, $x_2$).
We define the cross-term kernel as the averaging kernel of the second variable with the same coefficients as for the averaging kernel of the main variable:
\begin{equation} \mathcal K^* = \sum_i c_i K^i_{x_2, x_1}\,. \end{equation}
This expresses the contribution of the second variable $\delta x_2$ to the right-hand side of Eq.~\eqref{eqn:sola-approx}.
In order to be able to recover $\delta x_1$, this contribution needs to be sufficiently small.
Therefore, the amplitude of the cross-term kernel should be minimized, similar to the aim of the averaging kernel being localized around a certain point.
The uncertainty $\varepsilon$ on the estimate of $\delta x(r_0)$ would ideally also be as small as possible.
This uncertainty is determined by the coefficients $c_i$ and the covariance of all the observations: $\varepsilon = \sum_{ij} c_i c_j \Cov[\Delta_i, \Delta_j]$.
Finally, we also need to ensure that the effect of the scaling of the star via $\delta q$ does not influence the result.
This is accomplished by minimizing $\sum_i c_i$, as in that case the contribution from the scaling cancels out.
We therefore need to minimize the following function instead of just the term in Eq.~\eqref{eqn:sola-simple}:
\begin{equation}
  \begin{aligned}
      \mathcal{L} &= \int_0^R \left(\mathcal{K} - \mathcal{T}\right)^2\dd{r} + \beta \int_0^R \left(\mathcal{K}^*\right)^2\dd{r} + \mu \varepsilon + \alpha \left(\sum_i c_i\right)^2\,,
  \end{aligned}
  \label{eqn:sola-minimization}
\end{equation}
where the parameters $\beta$, $\mu$, and $\alpha$ weigh the importance of the cross-term kernel $\mathcal{K}^*$, the observational uncertainties and the effect of the scaling parameter, similar to the regularization parameter for RLS.
The normalization of the averaging kernel is fixed with a Lagrange multiplier.

The advantage of using SOLA is that it is possible to control how much the inferred profile at a certain point is influenced by other parts of the star.
How much the leakage from other parts of the star (i.e. the sidelobes of the averaging kernel) is suppressed can then be balanced with other requirements through the parameters $\beta$, $\mu$, and $\alpha$.
Moreover, the averaging kernel only depends on the uncertainties on the observations and not the observations itself.
This makes it possible to reuse and test the averaging kernel with just models instead of the observations.
The coefficients $c_i$ and thus the averaging kernel are then kept the same when using the frequency differences from the actual observations, essentially developing an inversion method for that specific reference model and set of observations.

\section{Stellar model grid}
\label{sec:model}

We use a model grid computed with MESA r15140 \citep{paxton-modules-2011,paxton-modules-2013,paxton-modules-2015,paxton-modules-2018,paxton-modules-2019} and GYRE 6.0.1 \citep{townsend-gyre-2013,townsend-angular-2018,goldstein-contour-2020}, with model parameters inspired by the forward modeling of KIC~10526294 \citep{papics-kic-2014,moravveji-tight-2015,triana-internal-2015}.
This star is one of the few SPB stars rotating at a slow enough rate such that the Coriolis acceleration can be ignored in the computation of the oscillation modes \citep{aerts-rossby-2021}.
As this star is also the only SPB pulsator with cleanly identified frequency triplets, it is a promising first candidate for future g-mode based inversions.
The parameter ranges (see Table~\ref{table:model-grid}) for our model grid are chosen to roughly correspond to the uncertainties from forward modeling of KIC~10526294 by \citet{moravveji-tight-2015}.
The shape of the mixing profile near the convective boundary is given by the extended convective penetration model with additional mixing in the envelope introduced by \citet{michielsen-probing-2021}.
In order to sample most of the parameter space, we randomly choose 192 sets of the initial mass, metallicity and three mixing parameters ($f_\mathrm{CBM}, \alpha_\mathrm{CBM}$ and $\log D_\mathrm{env}$) from the ranges in Table~\ref{table:model-grid}.
We fix $f_0$ to $0.005$.
Following \citet[][see their Section~4.1]{michielsen-probing-2021}, we use the OB star mixture deduced by \citet{nieva-presentday-2012} and \citet{przybilla-hot-2013} with the OP opacity tables \citep{seaton-opacity-2005}.
The initial helium fraction is computed using the enrichment law $Y_{\rm ini} = Y_p + 2.1\cdot Z_{\rm ini}$, with the primordial helium fraction set to $Y_p = 0.2465$ \citep{aver-primordial-2013}.
The factor 2.1 was chosen by \citet{michielsen-probing-2021} to match the chemical mixture from \citet{nieva-presentday-2012}.
For consistency, the initial hydrogen fraction is set to $X_{\rm ini} = 1 - Y_{\rm ini} - Z_{\rm ini}$.
For each set of input parameters, we compute evolutionary tracks with MESA and save models at intervals of $0.005$ in core hydrogen fraction between $X_c = 0.64$ and $X_c = 0.57$, for a total of 2880 stellar models.
This set of models is used as target models, while we take a model that lies close to center of the parameter space as a reference model, with parameters given in Table~\ref{table:model-parameters}.
We also highlight four different target models that we use as examples in the next sections.\
\footnote{The MESA and GYRE inlists and inversion software developed for this project are available on \url{https://doi.org/10.5281/zenodo.7941755}.}

\begin{table}
  \caption{Parameter ranges for the model grid}
  \label{table:model-grid}
  \centering
  \begin{tabular}{c l  l}
    \hline
    \hline
    Parameter & Lower limit & Upper limit \\
    \hline
    $M$ & 3.0 M$_\odot$ & 3.4 M$_\odot$ \\
    $Z$ & 0.012 & 0.028 \\
    $f_{\rm CBM}$ & 0.0 & 0.025 \\
    $\alpha_{\rm CBM}$ & 0.0 & 0.3 \\
    $\log D_{\rm env}$ & 1.0 & 2.0 \\
    $X_c$ & 0.57 & 0.64 \\
    \hline
  \end{tabular}
  \tablefoot{The parameters in this table are: the initial mass ($M$) and metallicity ($Z$), the mixing parameters ($f_\mathrm{CBM}, \alpha_\mathrm{CBM}, \log D_{\rm env}$), and the remaining core hydrogen fraction ($X_c$). $D_{\rm env}$ is expressed in cm$^2$s$^{-1}$ with $\log$ the base 10 logarithm.}
\end{table}

\begin{table}
  \tabcolsep=2.5pt
  \caption{Parameters of the reference model and the highlighted target models}
  \label{table:model-parameters}
  \begin{tabular}{l l l l l l l l}
    \hline
    \hline
    & $M$ & $Z$ & $f_{\rm CBM}$ & $\alpha_{\rm CBM}$ &  $\log D_{\rm env}$ & $X_c$ & $q / q_{\rm ref}$ \\
    & [M$_\odot$] & & & & & & \\
    \hline
    Reference  & 3.20 & 0.020 & 0.015 & 0.15 & 1.5 & 0.60 & \\
    Target \#1 & 3.15 & 0.019 & 0.023 & 0.06 & 1.1 & 0.60 & 0.988\\
    Target \#2 & 3.13 & 0.014 & 0.010 & 0.06 & 1.5 & 0.57 & 0.984\\
    Target \#3 & 3.36 & 0.014 & 0.023 & 0.05 & 1.7 & 0.58 & 1.005\\
    Target \#4 & 3.11 & 0.016 & 0.006 & 0.23 & 1.9 & 0.60 & 0.958\\
    \hline
  \end{tabular}
\end{table}

\begin{figure}
  \resizebox{\hsize}{!}{\includegraphics{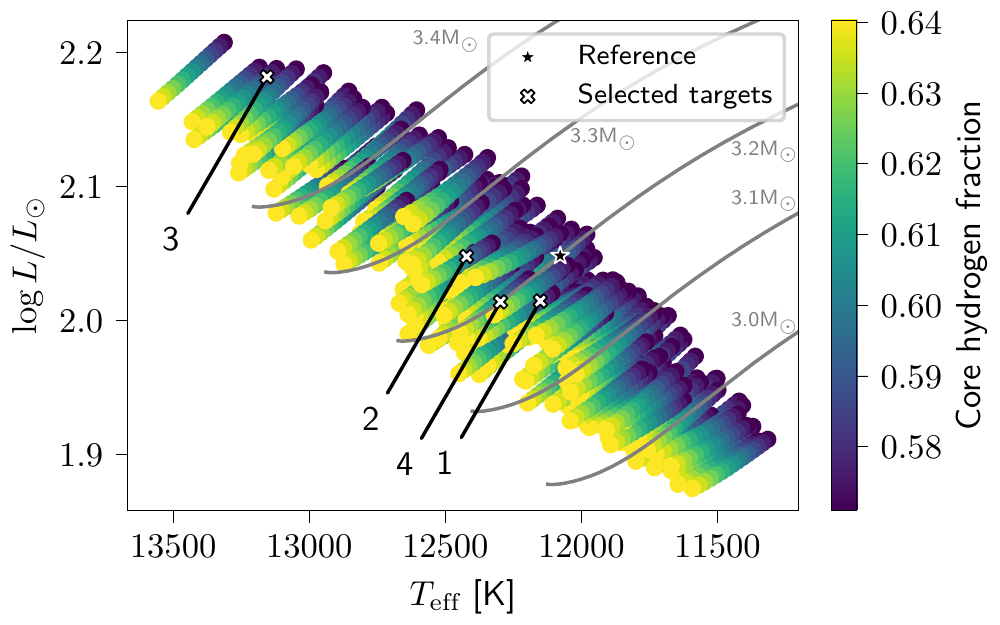}}
  \caption{Location of the evolutionary models in an HRD. Each colored circle is a stellar model, with the color indicating the hydrogen fraction in the core of the model. The reference model used throughout this work is indicated by a black star, while the white crosses are the target models. We refer to the target models by the numbers given in this plot. The gray lines show evolutionary tracks for different masses, with the same metallicity and overshooting parameters as the reference model.}
  \label{fig:model-sample}
\end{figure}

\begin{figure}
  \resizebox{\hsize}{!}{\includegraphics{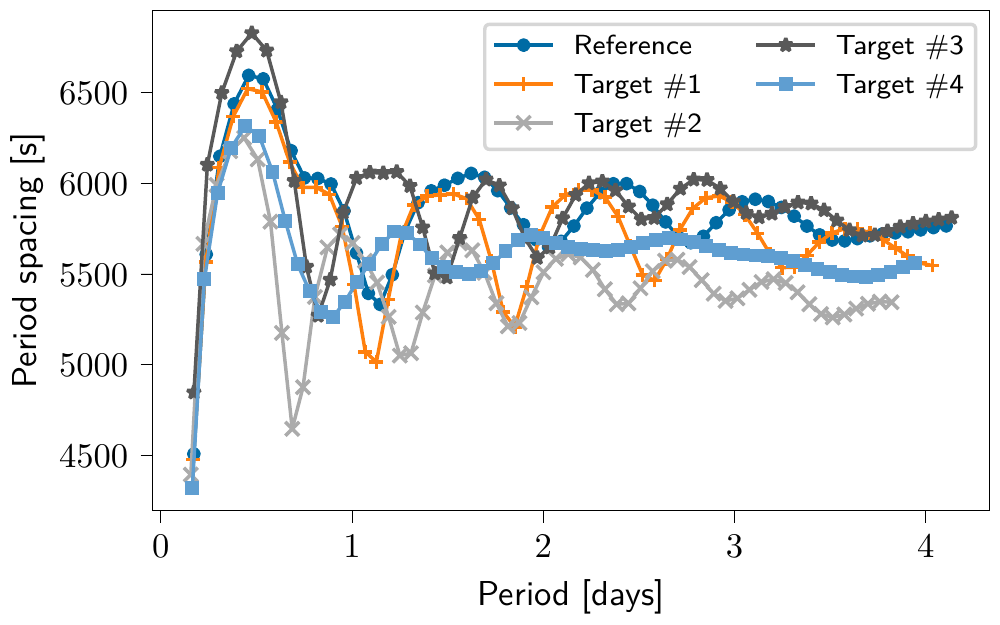}}
  \caption{Period-spacing patterns of the oscillation modes of the reference and target models. The oscillation modes considered are dipole g modes with radial orders from $n = 1$ to 60.}
  \label{fig:models-period-spacing-patterns}
\end{figure}

\begin{figure}
  \resizebox{\hsize}{!}{\includegraphics{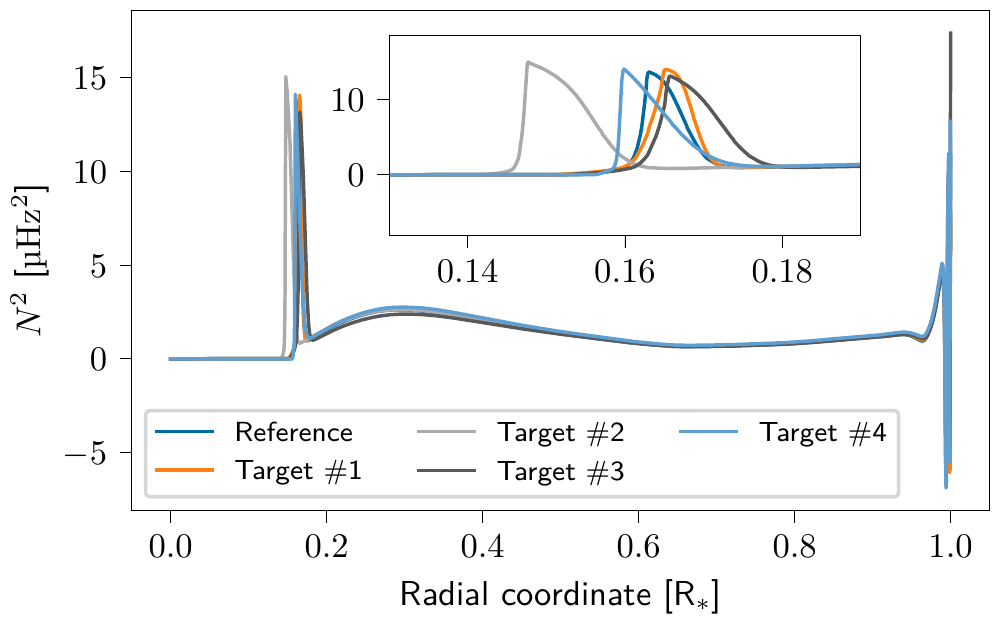}}
  \caption{Squared buoyancy frequency of the different target models. The peaks around \SI{0.16}{\radius\stellar} show where the transition from the envelope of the star to the core takes place and is the dominant factor in determining the period-spacing pattern.}
  \label{fig:models-buoyancy}
\end{figure}

The positions of the models in the Hertzsprung-Russell diagram (HRD), including the reference model and labeled target models, are shown in Fig.~\ref{fig:model-sample}.
Many of the tracks overlap due to the 6-dimensional nature of the modeling space.
Figure~\ref{fig:models-period-spacing-patterns} shows the dipole-mode period-spacing patterns of the reference model and the four highlighted target models.
The dip structure in the period spacing is well known for SPB pulsators \citep{miglio-probing-2008}.
It is caused by mode trapping within the near-core peak of the buoyancy frequency \citep{michielsen-probing-2019,michielsen-probing-2021}.
The position, extent, and shape of the peak in the buoyancy frequency determines where the dips in the period spacing pattern fall and how deep they are.
We explore this effect in the next section.
The corresponding buoyancy frequency profiles of these models can be seen in Fig.~\ref{fig:models-buoyancy}.
These profiles can essentially be split in two parts: a peak just above the convective core due to the chemical gradient left behind by the contraction of the core, and a slowly varying region in the envelope of the star determined by the temperature gradient.
At the surface of the star, atmospheric effects cause the buoyancy frequency to vary rapidly, but this is not relevant for g modes as they mostly probe the inner regions of the envelope.
The height of the near-core peaks of these profiles is similar to those of the best forward model solutions found in \citet{moravveji-tight-2015}.

\section{Nonlinear structure dependencies}
\label{sec:avoided-crossings}

The main assumption used in deriving the inversion equations in Sect.~\ref{sec:inversion} is that of linearity.
Both the quality of fit and the oscillation properties of the forward and reference model determine the magnitude of the error induced by this assumption.
How this error compares to other sources of uncertainty is explored in the next section.
In this section we explore the main source of nonlinear errors: the presence of avoided crossings or mode bumpings.
In the case of evolved solar-like pulsators, frequencies of p and g modes may occur in a similar range and can appear to cross or bump into each other as the star evolves.
Depending on the degree and the azimuthal order of the mode, instead of actually crossing, the frequencies of the modes may ``bounce off each other.''
This is not a physical interaction, as the oscillation equations are still linear, but rather the mathematical result of the coupling of the g- and p-mode cavities.
Avoided crossings have extensively been studied in the context of p/g-mixed modes \citep[e.g.,][]{scuflaire-non-1974,osaki-nonradial-1975,aizenman-avoided-1977,dziembowski-potential-1991,miglio-gravity-2008,ong-semianalytic-2020}, but lately also for couplings between core inertial modes and envelope gravito-inertial modes in fast rotators \citep{ouazzani-first-2020,saio-rotation-2021,lee-overstable-2021,tokuno-asteroseismology-2022}.

\begin{figure}
  \resizebox{\hsize}{!}{\includegraphics{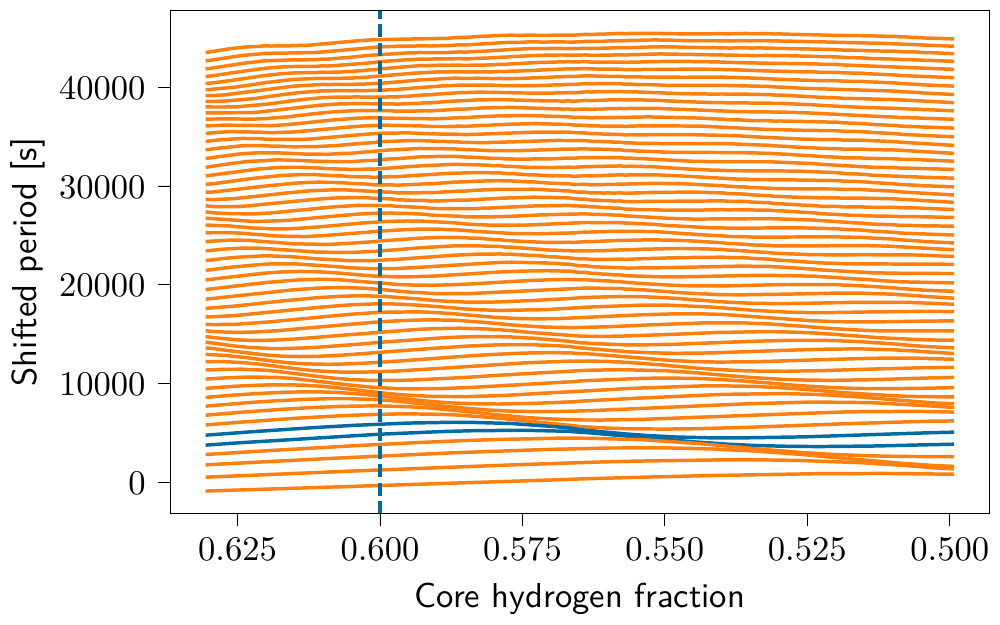}}
  \caption{Evolution of the periods of the oscillation modes of the evolutionary track containing the reference model. The periods have been shifted by $-n\SI{5000}{\second}$, where $n$ is the radial order of the mode, so that the change in mode periods is more pronounced. Radial orders $n = 6$ to $n = 60$ (from bottom to top) have been plotted. The two modes highlighted in blue are the $n = 10$ and $n = 11$ modes shown in Fig.~\ref{fig:avoided-crossings-displacement-compare}. The denser regions show where the dips in the period-spacing patterns are and how they evolve as the star ages. Where these avoided crossings are located is strongly dependent on the size and location of the buoyancy frequency peak and we can therefore learn about the near-core mixing processes from them. The vertical dashed line indicates the reference model.}
  \label{fig:avoided-crossings}
\end{figure}

\begin{figure}
  \resizebox{\hsize}{!}{\includegraphics{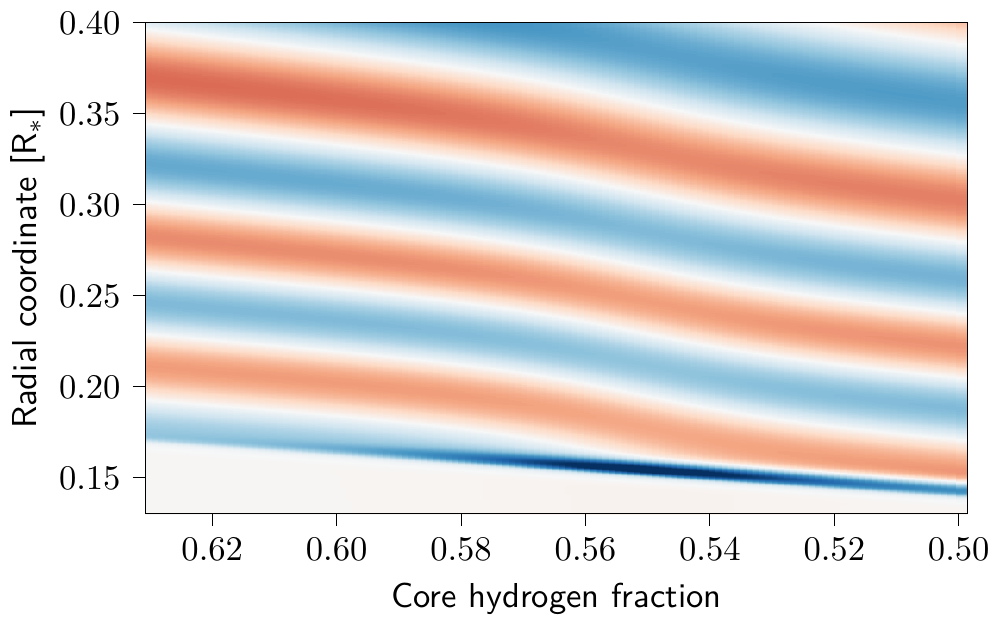}}
  \caption{Evolution of the horizontal displacement of the $n = 10$ g mode of the evolutionary track containing the reference model. The colors and intensity indicate the sign and amplitude of the displacement. Since the oscillation equations are linear and homogeneous, the absolute magnitude and overall sign are not relevant. Around $X_c = 0.56$ the mode undergoes avoided crossings with the $n=11$ and $n=9$ modes, causing strong changes in the structure of the mode.}
  \label{fig:avoided-crossings-displacement}
\end{figure}

\begin{figure}
  \resizebox{\hsize}{!}{\includegraphics{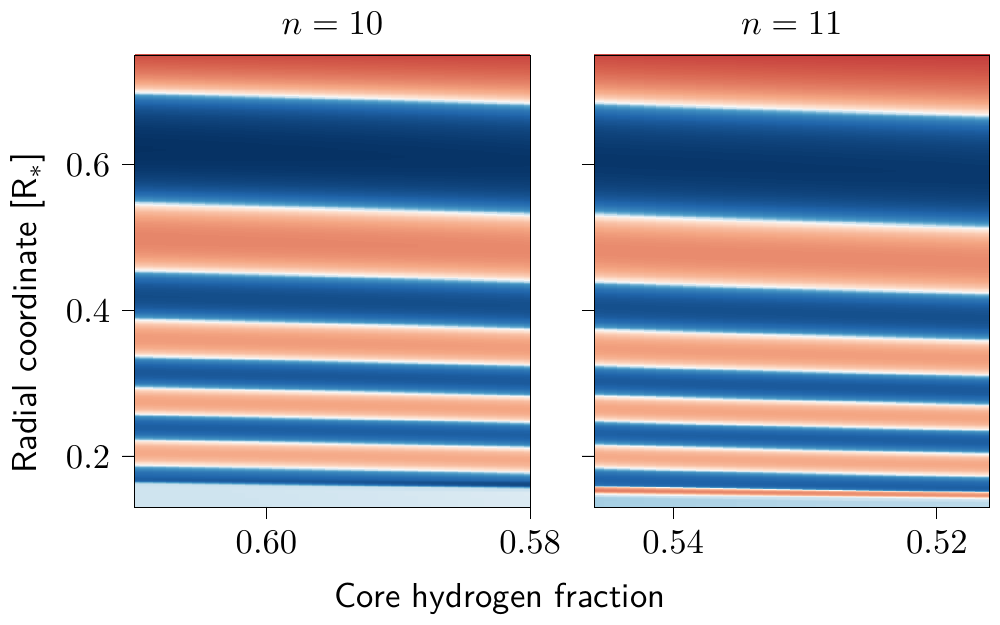}}
  \caption{Slices of the evolution of the $n = 10$ and $n = 11$ g modes of the reference model, as in Fig.~\ref{fig:avoided-crossings-displacement}, but now plotted with logarithmic amplitude. The left panel shows the $n = 10$ mode before its avoided crossing with the $n = 11$ mode, while the right panel shows the $n=11$ mode after its avoided crossing with the $n = 10$ mode. This shows that the structure of the $n = 10$ mode before the crossing is similar to the $n = 11$ mode after the crossing, with the exception of an additional node near the core of the star.}
  \label{fig:avoided-crossings-displacement-compare}
\end{figure}

\begin{figure}
  \resizebox{\hsize}{!}{\includegraphics{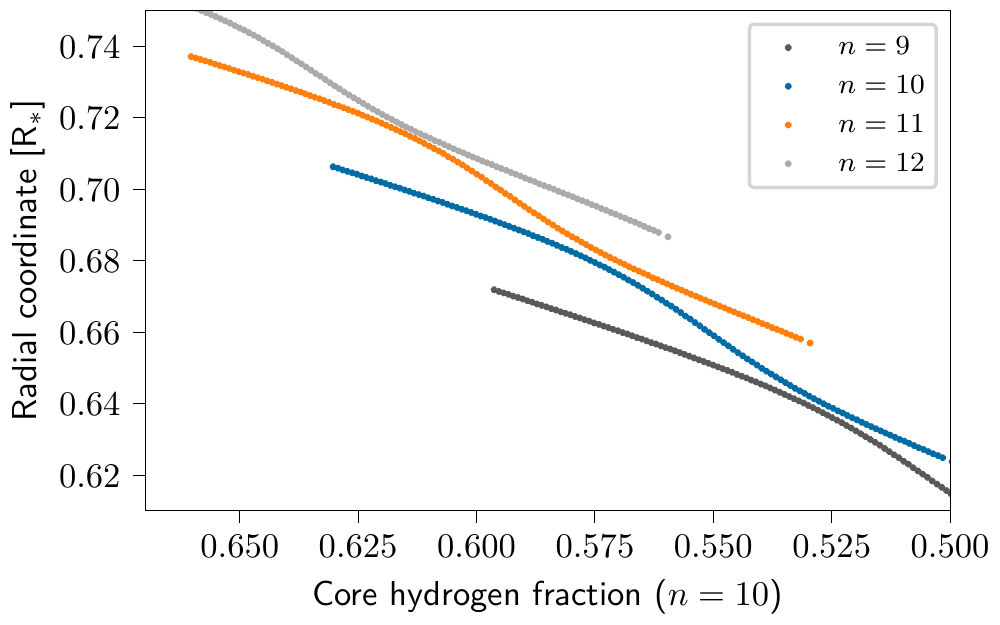}}
  \caption{Evolution of the location of the nodes closest to the surface of the $n=9, 10, 11$ and 12 modes. In order to increase visual clarity, the nodes of the $n=9, 11$ and 12 modes have been shifted by values of $0.034, -0.030$ and -0.060 in core hydrogen fraction. This aligns the modes before and after the avoided crossing, as in Fig.~\ref{fig:avoided-crossings-displacement-compare}. Immediately after the first cascade of avoided crossings, the $n = 12$ mode has settled in at the position of the $n=11$ mode just before the cascade, the $n=11$ mode at the position of the $n=10$ mode and finally the $n=10$ mode at the position of the $n=9$ mode. }
  \label{fig:avoided-crossings-top-node}
\end{figure}

\begin{figure}
  \resizebox{\hsize}{!}{\includegraphics{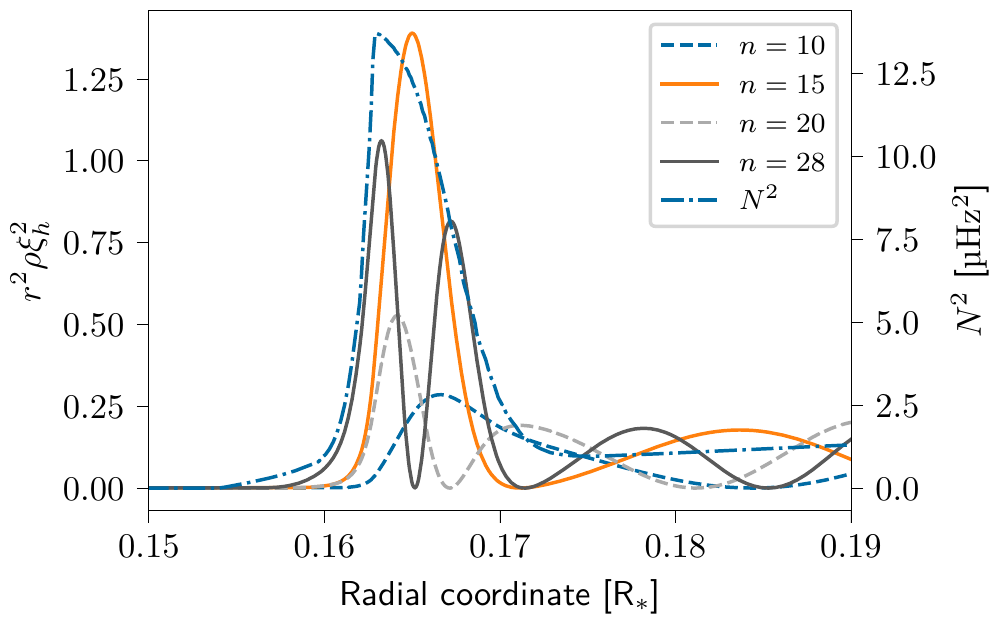}}
  \caption{Square of the horizontal displacement, scaled by $r^2 \rho$, for dipole modes with indicated radial order of the reference model. This reflects the energy density of the mode per unit radial distance, as a rough measure of the sensitivity of the mode to the structure. The modes which are in the dip of the period-spacing pattern (solid lines) have a much higher amplitude in the peak of the buoyancy frequency than those not in the dips of the period-spacing pattern (dashed lines). The squared buoyancy frequency is shown as a blue dash-dotted line.}
  \label{fig:resonance}
\end{figure}

The dips in the period spacing pattern we discussed in the previous section have a similar origin.
Figure~\ref{fig:avoided-crossings} shows the evolution of the oscillation periods of the reference model track.
As the star ages, a cascade of avoided crossings moves downward in period.
During such a cascade, the periods of trapped modes with nearby radial orders are brought close together, causing the dips in the period-spacing patterns.
Each mode will undergo two avoided crossings during such a cascade, one with each of the two modes adjacent in radial order.
The changes in mode period during these avoided crossing are quite small, on the order of 10\% of the spacing between the periods, compared to for example the p/g-mode avoided crossings in evolved stars where one of the oscillation modes essentially ends up at the frequency of the other mode \citep[see Fig. 4 of][]{bellinger-asteroseismic-2021}.
When a g mode undergoes an avoided crossing, not only does its period change rapidly compared to the structure changes, its eigenfunction will also change significantly.
Figure~\ref{fig:avoided-crossings-displacement} shows how the horizontal displacement of the $n =10$ dipole mode of the reference model evolves as the star ages.
During the avoided crossing, which happens around $X_c = 0.56$, the radial nodes move down rapidly, with the displacement function peaking at the edge of the core.
After the avoided crossing, the displacement seems to behave more similar to the eigenfunction of the other mode in the avoided crossing, the $n = 11$ mode in this case, as can be seen in Fig.~\ref{fig:avoided-crossings-displacement-compare}.
The radial nodes of the $n = 11$ mode after the avoided crossing are located at and spaced similarly to the nodes of the $n = 10$ mode before the avoided crossing.
Figure~\ref{fig:avoided-crossings-top-node} shows how the node closest to the surface of the star evolves for the modes with $n = 9-12$.
It is clear that the position of the node as a function of the core hydrogen fraction is transferred between the different modes at the avoided crossings.
However, the exchange of properties is not exact, in the sense that the $n = 11$ mode still retains some of its own character.
Indeed, it remains possible to distinguish the $n = 10$ and $n = 11$ modes, as the number of radial nodes is always clearly visible.
Close to the core, the radial nodes are shifted in the direction of the surface due to the additional node in the peak of the buoyancy frequency.
A similar effect is found in p/g-mixed modes, where the interaction causes the p modes to gain a g-mode character and vice-versa while they retain some of their original p- or g-mode character \citep[e.g.,][]{smeyers-linear-2010}.
For the pure g-mode pulsators considered in this work, however, only one mode cavity is active.
Instead, the avoided crossings occur due to a resonance between the modes and the rapidly evolving peak in the buoyancy frequency at the shrinking convective core causing mode trapping.
Figure~\ref{fig:resonance} shows the horizontal displacement of different oscillation modes as a function of the radial coordinate.
For modes that are in an avoided crossing (the $n = 15$ mode), the mode energy is amplified by the peak of the buoyancy frequency, with avoided crossings at higher radial order behaving as overtones, showing multiple enhanced peaks.
Similar results were found earlier by \citet{michielsen-probing-2021}; readers can refer to their Fig.~1.

The main challenge of avoided crossings is that they cannot be described using a first-order formalism.
Since the inversion methods considered in this paper build upon the assumption of linearity, the presence of avoided crossings induces errors.
In the next sections we determine how large this error is in comparison with observational uncertainties for the mode frequencies, for both RLS and SOLA.
This helps us to determine whether structure inversions of main-sequence g-mode pulsators requires nonlinear inversions that take the numerous avoided crossings of SPB stars into account and whether such methods will yield improved inferred profiles given the observational uncertainties.

\section{Testing the inversion methods}
\label{sec:test}
\subsection{Assessing the quality of the kernels}
\label{sec:predictions}

\begin{figure}
  \resizebox{\hsize}{!}{\includegraphics{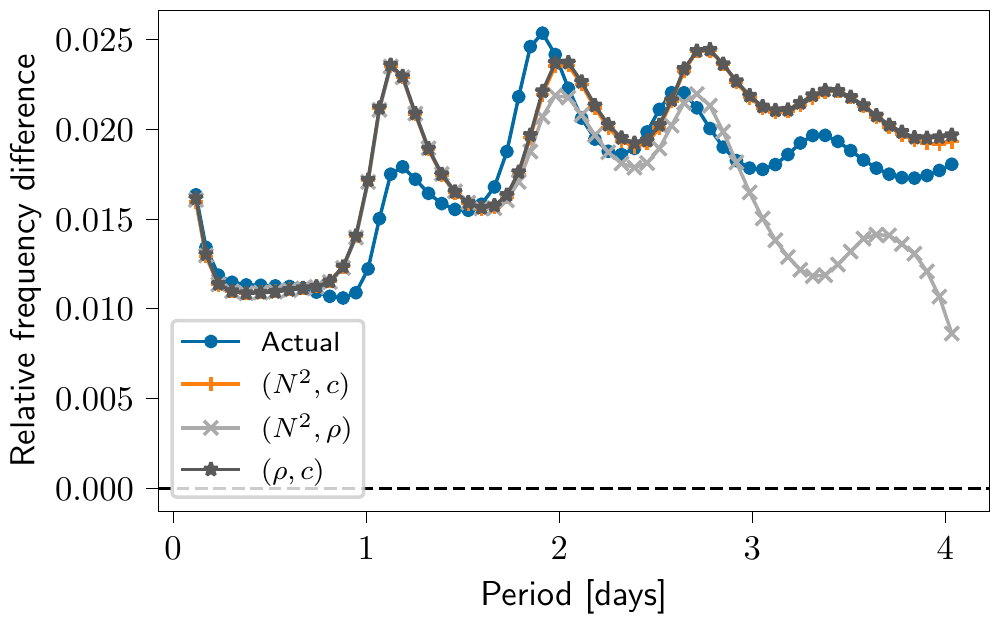}}
  \caption{Predictions of the frequency differences between the reference model and the closest target model, target model \#1. The dark blue circles show the actual difference as computed using GYRE. The three other types of points are computed using the oscillation kernels for the variable pairs $(N^2, c)$, $(N^2, \rho)$ and $(\rho, c)$. While the $(\rho, c)$ and $(N^2, c)$ variable pairs agree with each other, the $(N^2, \rho)$ variable pair shows significant deviations.}
  \label{fig:model-1-variable-pairs}
\end{figure}

\begin{figure}
  \resizebox{\hsize}{!}{\includegraphics{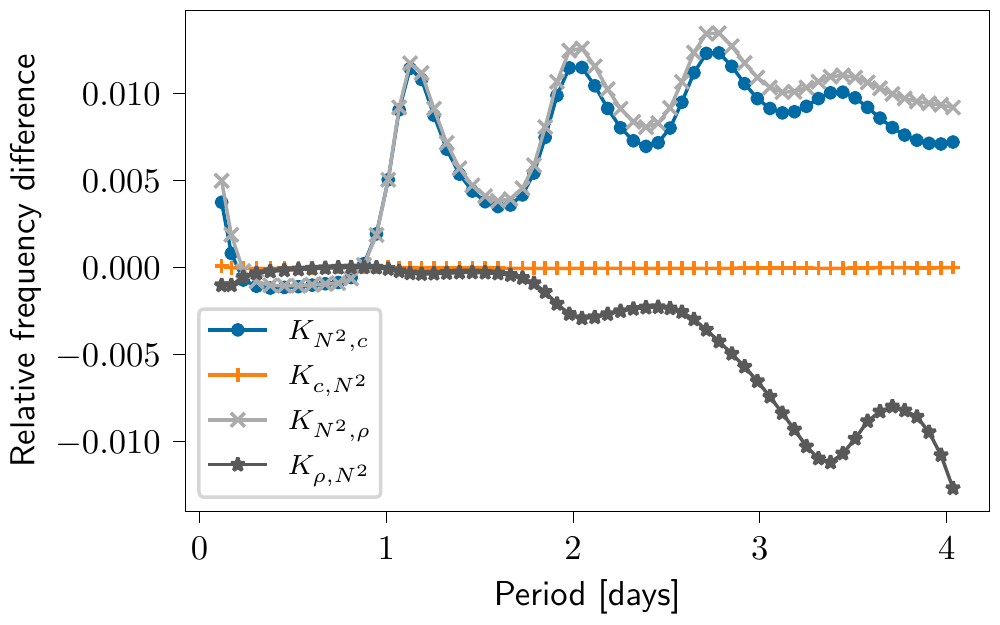}}
  \caption{Contributions of the individual variables to the predicted frequency differences in Fig.~\ref{fig:model-1-variable-pairs}. Only the variable pairs $(N^2, c)$ and $(N^2, \rho)$ are shown here. Adding the contributions for the two variables in a variable pair does not result in the values shown in Fig.~\ref{fig:model-1-variable-pairs} as the $\delta q / q$ term needs to be added as well (see Eq.~\ref{eqn:kernel-definition}).}
  \label{fig:model-1-variables}
\end{figure}

In order to determine the quality of the oscillation mode kernels, we compare the frequencies of the target models computed with GYRE to the frequencies predicted from the reference model using Eq.~\eqref{eqn:kernel-definition}.
This not only allows us to understand how well the linear approximation holds for a given reference and target model, but also allows us to identify which pair of variables is optimal for inversions.
Fig.~\ref{fig:model-1-variable-pairs} shows both of these elements for target model \#1.
While the predictions do seem to match the overall properties of the frequency differences between the two models, they already show a significant error (up to 25\% of the actual difference, for a total of $\sim0.5$\%).
This error is larger than typical uncertainties on observed frequencies, which have an order of magnitude of 0.01\% \citep{vanbeeck-detection-2021}.
All of the variable pairs show similar predictions of the oscillation frequencies, with the $(c, \rho)$ and $(N^2, c)$ pairs matching, while the $(N^2, \rho)$ pair deviates at higher radial orders.
The reason for this is unclear, but could possibly be due to numerical errors in the computations of the kernels (see Appendix~\ref{sec:kernel-derivations}).
The same predictions as in Fig.~\ref{fig:model-1-variable-pairs} can be seen in Fig.~\ref{fig:model-1-variables} but for each of the variables of the $(N^2,\rho)$ and $(N^2, c)$ pairs separately.
For the $(N^2, \rho)$ variable pair, both variables contribute about equally to the total difference, although this depends a bit on the radial order.
In the case of the $(N^2, c)$ variable pair, however, the contribution of the sound speed is completely suppressed compared to the contribution of the buoyancy frequency.
Hence, for the next steps we only use the $K_{N^2, c}$ kernel for both the predictions and the inversions.
We note that since we could describe the change in frequencies with only one variable, trying to invert the structure for two variables would introduce degeneracies between those two variables.

\begin{figure}
  \resizebox{\hsize}{!}{\includegraphics{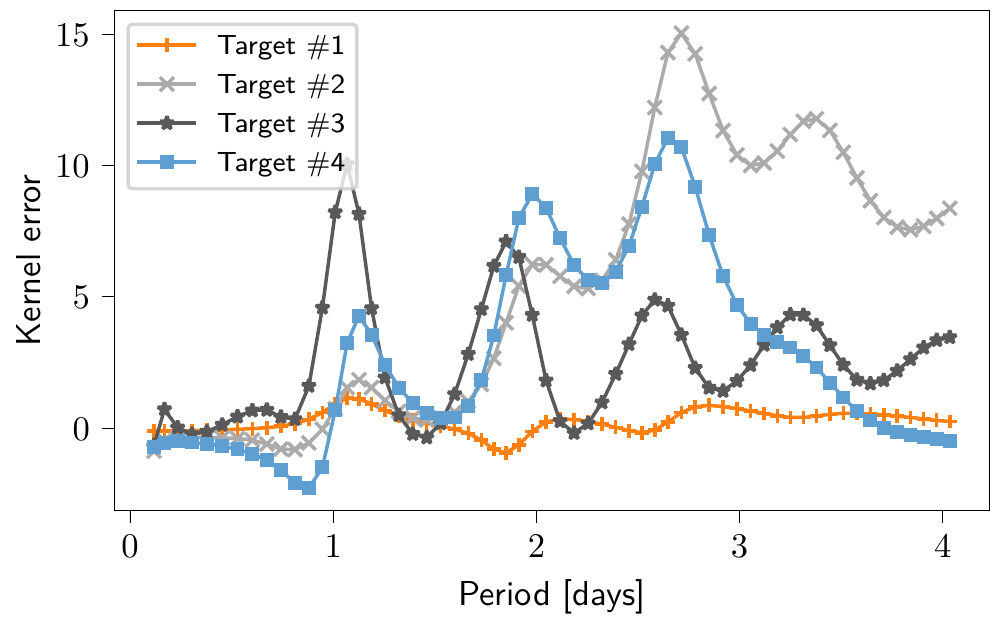}}
  \caption{Error of the predicted frequency differences using the $(N^2, c)$ variable pair, scaled by the average absolute frequency difference between the reference and target models after removing the effect of the scale change $\delta q$. An error larger than 1 therefore means that the error is larger than the overall difference between the models. For target model \#1, this is essentially Fig.~\ref{fig:model-1-variable-pairs} divided by the average difference between the models.}
  \label{fig:model-estimation}
\end{figure}

Moving on to the other target models with larger differences with the reference model (see Fig.~\ref{fig:model-estimation}), we find that predictions of the target model's frequencies have large errors compared to the difference in frequency.
The size of this error depends on the radial order of the oscillation mode, and the peaks in the error roughly coincide with the dips in the period-spacing patterns in Fig.~\ref{fig:models-period-spacing-patterns}.
This indicates that these errors are primarily caused by the avoided crossings in these models, as they are the cause of the dips in the period spacing pattern of the reference model.
The exact location of the peaks in the kernel error differs between the target models as the changes in the structure also change which modes are in avoided crossings.
The lowest-order modes suffer significantly less from these errors compared to the higher-order ones at higher periods.
The overall magnitude of these errors is large, often even larger than the difference in frequencies to begin with.
However, as this error is significantly correlated between the modes and a considerable number of modes has a low error, this cannot be directly compared to other sources of uncertainty such as the observational errors.
Therefore, in the next two sections we consider how well RLS and SOLA are able to infer the correct structure of the target models from the reference model.
This allows us to identify whether the kernels provide enough information to recover the details of the structure of the star.

\subsection{RLS}

The first inversion method, RLS, requires two additional input parameters aside from the observations and the reference model.
A choice for the regularization parameter needs to be made, manually or via methods such as cross-validation or the L-curve \citep{hansen-lcurve-2001}, as well as how the radial coordinate is discretized.
For the latter choice, not only is it relevant how dense the resulting grid is, but also whether it is uniform or not.
Due to the irregular shape of the buoyancy frequency, the oscillation kernels do not probe each region of the star equally, but rather predominantly probe the small near-core region, where the peak in the buoyancy frequency is located.
In the region where the probing power of the g modes is higher, more freedom should be given to the inferred profile, while less so in other parts of the star.
We use a step function as the inferred profile, with the spacing between the grid points depending on the location in the star.
In the near-core region, where the peak in the buoyancy frequency is located, we use a small spacing between the grid points, whereas in the other parts of the star, we use a larger spacing.
In each region the spacing is constant.
How refined the grid needs to be in the near-core region and what the boundaries of this region should be, needs to be determined in a similar way to the regularization parameter.
In our first attempt at g-mode inversions using RLS, we manually tune the parameters based on the actual difference between the models.
If these parameters are consistent between the different targets models, they can be reused for inverting actual observations given the same reference model \citep[e.g.,][]{christensen-dalsgaard-comparison-1990,deheuvels-seismic-2012}.
We fix the observational uncertainty on the oscillation frequencies at $10^{-3}$\SI{}{\micro\hertz} as an order-of-magnitude estimation of actual observational uncertainties.
This is on the order of 0.01\% relative uncertainty for an oscillation frequency of \SI{1}{\per\day}, which is typical for main-sequence g-mode pulsators \citep{aerts-angular-2019,vanbeeck-detection-2021}.

\begin{figure*}
  \resizebox{\hsize}{!}{\includegraphics{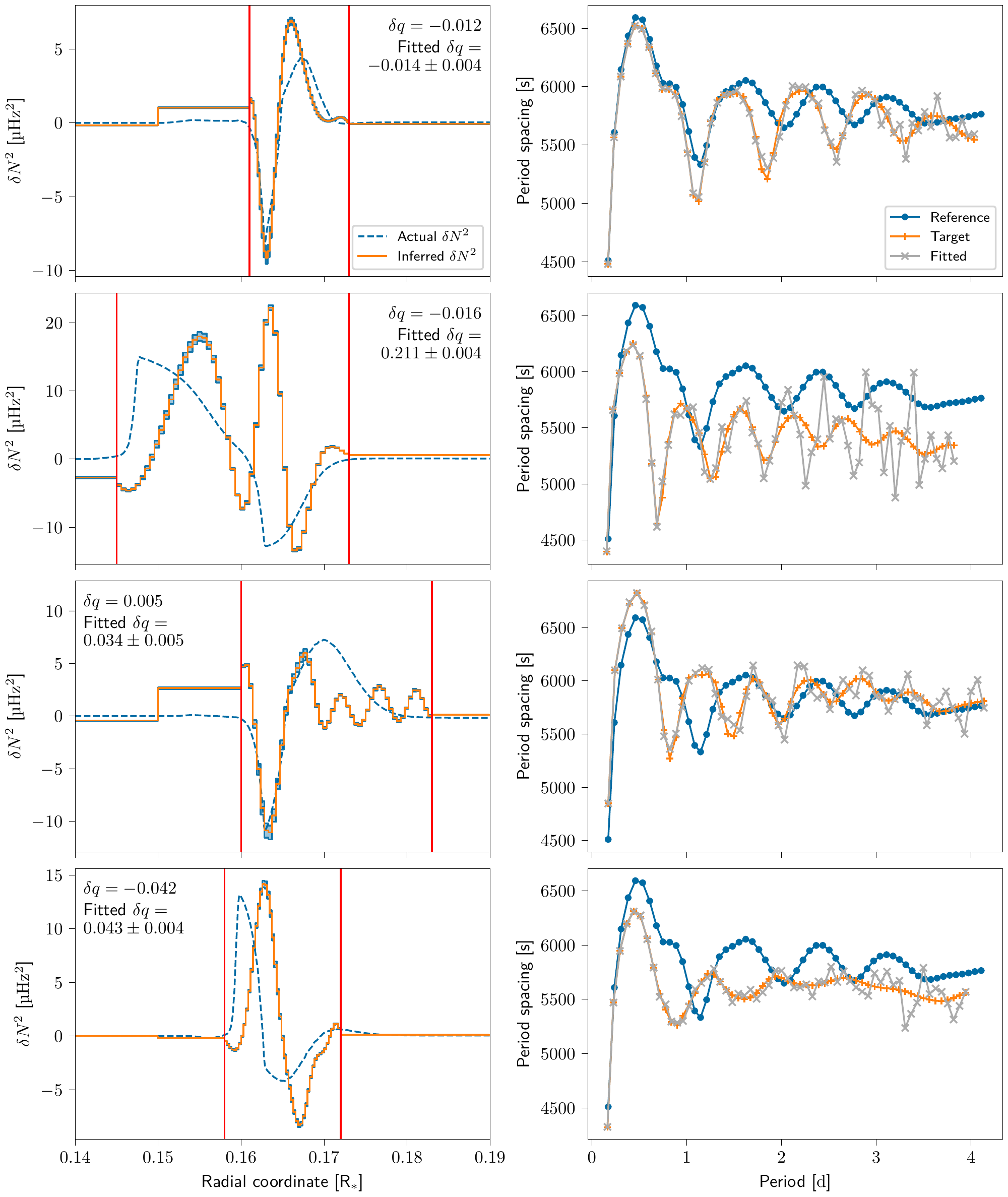}}
  \caption{Structure inversions of the four target models using RLS. The left column shows the inferred buoyancy frequency difference (orange full line) and the difference between the models (blue dashed line) for target models \#1 to \#4 in order. The observational uncertainty is indicated by the light blue regions around the inferred profile. For most of these inferred profiles, the regions are approximately the size of the line width. The right column shows the period-spacing pattern of the reference model (blue circles), the target models (orange plus symbols) and the fitted modes (gray crosses; $\bar \Delta_i$ defined in Eq.~\ref{eqn:delta-bar}). The inferred profile is discretized in 70 grid points, of which 50 are located in regions of the largest difference in buoyancy frequency, indicated by the red lines. The regularization parameter varies between $10^{-4.8}$ and $10^{-4.2}$ between the different target models. Except for the closest target model, the inferred profiles show significant deviation from the actual difference between the models. The period-spacing patterns show that the overall quality of the fit is limited, especially at higher radial orders.}
  \label{fig:rls-4-targets}
\end{figure*}

\begin{figure*}
  \resizebox{\hsize}{!}{\includegraphics{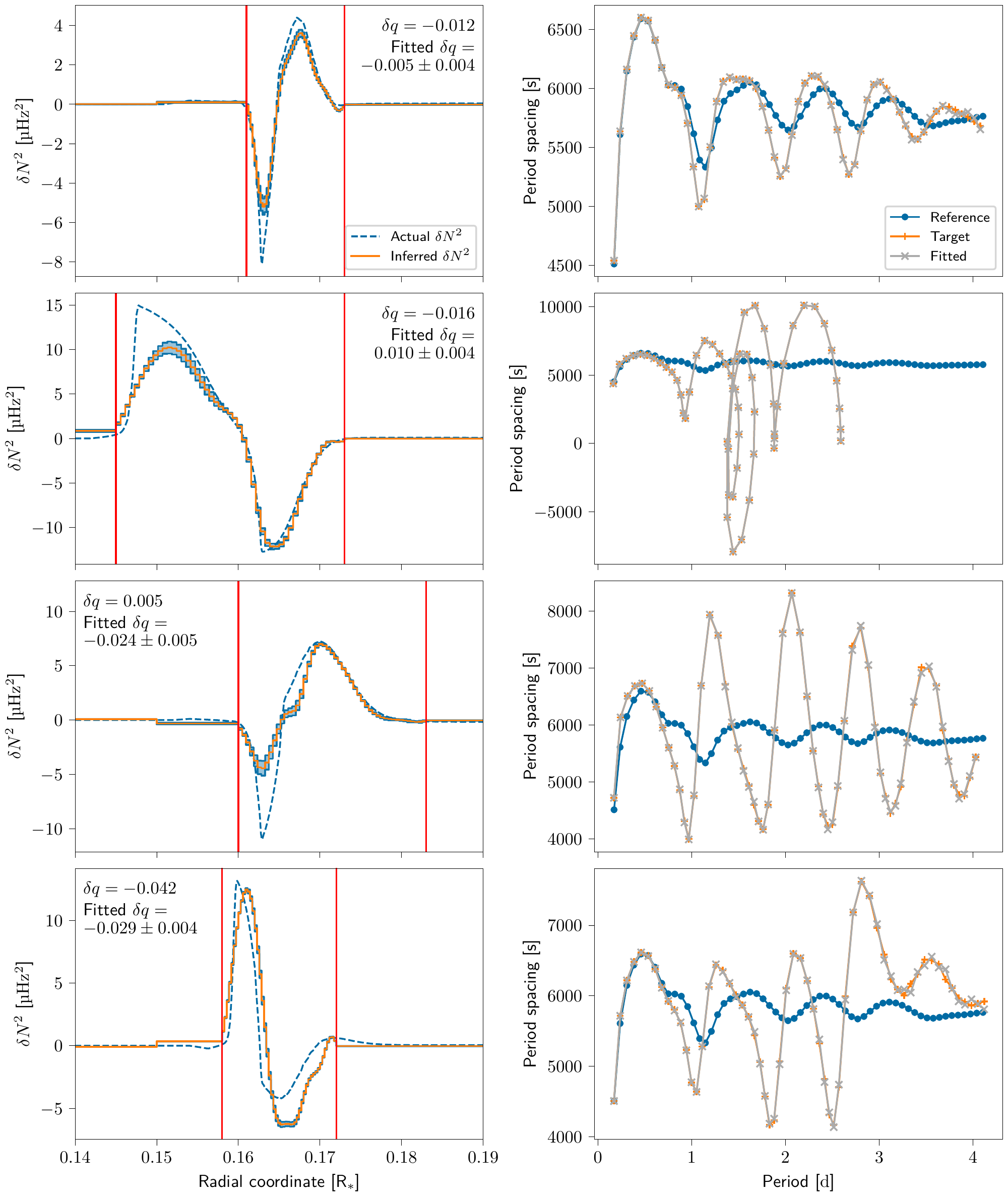}}
  \caption{Inversion of the same target models as in Fig.~\ref{fig:rls-4-targets}, but with the difference in oscillation frequencies calculated from the difference in structure and the oscillation kernels. This essentially performs the inversion as if there were no nonlinear dependencies on the structure. All inversion parameters were kept the same. The quality of the fit is significantly better and the inferred profiles match the actual difference between the models much better. The corrected period-spacing patterns show how strong the nonlinear dependency on the structure is, deviating significantly from the actual period-spacing patterns of the target models shown in Fig.~\ref{fig:rls-4-targets}. In the case of target model \#2 this even causes the oscillation modes to cross in frequency.}
  \label{fig:rls-4-targets-corrected}
\end{figure*}

Figure~\ref{fig:rls-4-targets} shows what would be considered the best possible case for an inversion for the buoyancy frequency of the different target models, given a full set of 60 modes.
However, the inferred difference is only close to the actual difference for target model \#1, while the profiles for the other targets show spurious features that are not present in the actual difference between the models.
If the refined region of the grid is not placed directly on the peaks in the difference between the models, the inversion does not successfully recover the difference.
Similarly, the valid ranges of regularization parameters that still lead to reasonable inferred profiles do not overlap between the target models.
Therefore, no single set of inversion parameters exists that is applicable to a wide range of potential real profiles.
As we need to tune these parameters manually, we did not apply the RLS inversion method in its linear approximation to the full stellar model grid.
Indeed, the automated methods mentioned above for determining these parameters did not provide satisfactory results.
Future nonlinear treatment of the structure dependency may make these methods meaningful again.
The fit also shows significant deviations for higher order modes as a result of the suppression of the second derivative.
The uncertainties from the observations are on average at least an order of magnitude smaller than the difference between the inferred and the actual profiles.
Hence the observational input is not be the main factor limiting inversions for this set of frequencies.

There are two potential reasons for the poor fit quality: the oscillation modes might not contain enough information about the details of the near-core region to properly resolve it, or the nonlinear dependencies on the structure dominate the fit.
To test which of these two reasons causes the poor fit, we remove all nonlinear dependencies by using the predicted frequencies from the previous section as the target frequencies rather than those computed with GYRE.
If the fit improves significantly, then the cause of the poor fit is the nonlinear effects.
Otherwise, the fit quality is most likely dominated by the information that is available in the modes.
The result for the same four target models can be found in Fig.~\ref{fig:rls-4-targets-corrected}, showing that the fit is of a much better quality.
For all four models the profile inferred from the inversion traces out the near-core region almost perfectly.
While not shown here, one can also find inversion parameters that are applicable to all target models, rather than having to tune the parameters to each target model.
In the core, on the other hand, the inferred profile shows significant deviation.
However, because the kernels only vanish completely at the center of the star, but are similar in shape and amplitude, this region is not constrained much, but does affect the frequencies and the fit.
The typical observational uncertainties on the inferred profile are still significantly smaller than the difference between the inferred and actual profiles.
Hence, the error in this case would be dominated by the lack of detailed information and the necessary regularization.
One final element of these fits is that the scaling of the star cannot be determined properly.
As is shown in Appendix\,\ref{sec:reduced-scaling}, most of the modes do not carry information about the scaling parameter.
Fixing the scaling parameter to the expected value does not change the inversion of the near-core region significantly, as long as the regularization parameter is not too large.

One of the potential ways to reduce the effect of the avoided crossings on the inferred profile, is to remove those modes that are nearest to the avoided crossings.
We explore this in Appendix~\ref{sec:change-number-freqs}.
We find that removing modes around avoided crossings does not improve the quality of the fit appreciably.
While the nonlinear effects are smaller when removing modes around the avoided crossings of the reference model, significant nonlinearity is still present in the other modes.
This is because the range of modes that are affected by avoided crossings changes quite fast for relatively small changes in the parameter space.
This could already be seen in Fig.~\ref{fig:avoided-crossings} for the core hydrogen fraction.

\subsection{SOLA}

The main task in performing an inversion with SOLA is deducing the optimal set of parameters used to determine the averaging kernel.
Not only do we need to determine where the averaging kernel can be localized and how wide it should be in that location, but we also need to tune the scaling and uncertainty suppression parameters (the parameters $\alpha$ and $\mu$ respectively in Eq.~\eqref{eqn:sola-minimization}).
As the effect of the $K_{c, N^2}$ kernel was shown to be negligible in Sect.~\ref{sec:predictions}, we set the parameter $\beta$ to zero, removing the contribution of the cross-term to Eq.~\eqref{eqn:sola-minimization}.
As for the case of RLS, we fix the observational uncertainties on the frequencies at $10^{-3}$\SI{}{\micro\hertz}.
We also consider an upper limit on the uncertainty caused by the scaling factor, by assuming that $\left|\delta q / q\right| \leq 0.2$, which is a conservative estimate of the uncertainty on stellar masses from forward modeling \citep{pedersen-diversity-2022}.
We do not use the exact same expression as was given in Eq.~\eqref{eqn:kernel-definition} to compute this uncertainty, but rather use the modified expressions from Appendix~\ref{sec:reduced-scaling} which reduce the effect of the scaling parameter by inverting for the dimensional buoyancy frequency.
The suppression parameters are then determined by choosing a desired upper limit on the uncertainty of the squared buoyancy frequency, which we for these testing purposes place at \SI{0.1}{\micro\hertz\squared}.
The contributions from the uncertainties on the frequencies and the scaling parameter are balanced, to prevent needlessly suppressing one of the contributions while the other remains large.

\begin{figure}
  \resizebox{\hsize}{!}{\includegraphics{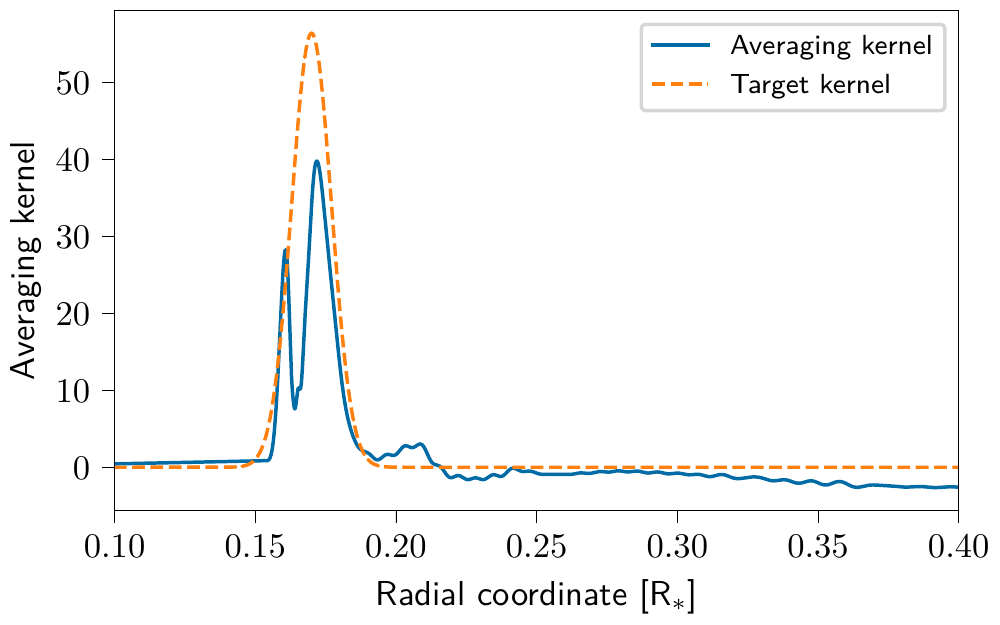}}
  \caption{Target and averaging kernel for the reference model, localized in the near-core region, based on dipole oscillation modes with radial orders $n=1$ to 60. The averaging kernels shows a double peak, which is located at the edges of the peak in the buoyancy frequency. The target kernel used is a modified Gaussian located at $r = 0.17$ and a width of 0.01.}
  \label{fig:sola-near-core}
\end{figure}

\begin{figure}
  \resizebox{\hsize}{!}{\includegraphics{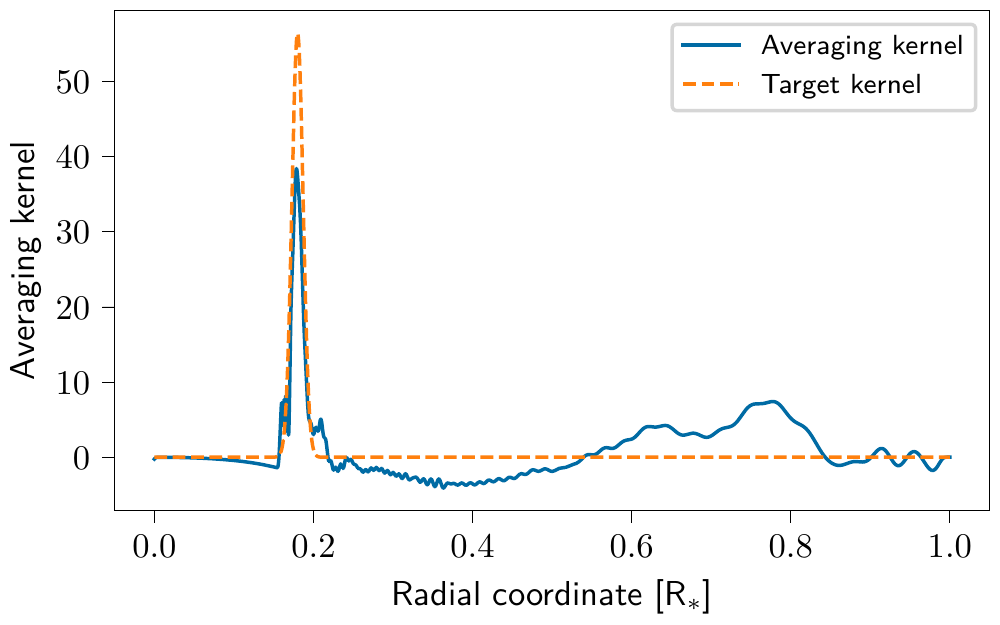}}
  \caption{Target and averaging kernel for the reference model, localized just above the near-core region, based on dipole oscillation modes with radial orders $n=1$ to 60. The averaging kernels shows a single peak at the location of the target kernel. Some residual amplitude remains in the envelope of the star. The target kernel used is a modified Gaussian located at $r = 0.18$ and a width of 0.01.}
  \label{fig:sola-outside-near-core}
\end{figure}

Based on manual experimentation with the width and location of the target kernel, we selected three distinct regions in the star: the near-core region (defined by the peak of the buoyancy frequency), just outside the near-core region and the envelope.
Each of these regions shows a different behavior for the averaging kernel.
We only show the results for a certain combination of oscillation frequencies, but it is important to keep in mind that the result will vary depending on which oscillation modes are actually observed and identified.
Starting with the near-core region, we find that the averaging kernel shows a bimodal behavior (see Fig.~\ref{fig:sola-near-core}) and cannot properly distinguish between changes at the lower and upper boundary of the peak in the buoyancy frequency.
Interestingly enough, the amplitude of the averaging kernel just around the peak is small, indicating that we have no localized information about the center of the peak, even when we choose the target kernel specifically in at this peak.
Changing the width or position of the target kernel does not eliminate this double-peaked feature, hence it seems to be an intrinsic feature of the oscillation modes.
For the regions just outside the near-core region (see Fig.~\ref{fig:sola-outside-near-core}), we find again that we are able to localize the averaging kernel, this time without the bimodal structure.
Finally, we find that the remaining parts of the envelope of the star cannot be resolved, as this always leads to a significant amplitude of the averaging kernel at other locations in the star.
This is expected, as it is well known that g modes have no probing power in the envelope.
In all cases, the suppression of the uncertainty on the inversions with these averaging kernels does not significantly affect the averaging kernels, showing that the uncertainties are not the limiting factor on what can be recovered from the oscillation frequencies.

\begin{figure}
  \resizebox{\hsize}{!}{\includegraphics{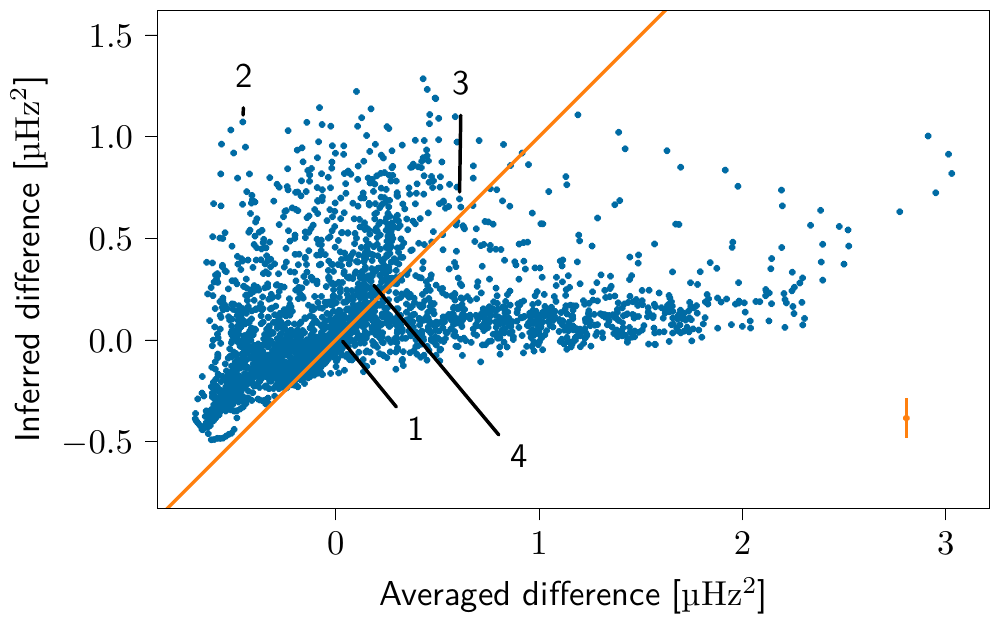}}
  \caption{Comparison of the actual difference in structure to the inferred difference for the averaging kernel in Fig.~\ref{fig:sola-outside-near-core}. The values on the $x$- and $y$-axis are differences in squared buoyancy frequency. The $x$-axis shows the integral of the structure difference multiplied by the averaging kernel (the integral in Eq.~\ref{eqn:sola-approx}), while the $y$-axis shows the inversion result (the left hand side of Eq.~\ref{eqn:sola-approx}). Each blue dot is one of the 2880 stellar models. The orange diagonal line is given by $x = y$. All deviations from this line are caused by the nonlinear dependency on the structure. The numbers indicate the different target models. The orange error bar in the lower right corner of the plot indicates the observational uncertainty on the inferred difference, given an uncertainty of $10^{-3}$\SI{}{\micro\hertz} on the observed frequencies.}
  \label{fig:sola-compare-models}
\end{figure}

Whether the averaging kernels can be localized and what the observational uncertainty on the actual result is, is only relevant if the kernels can accurately predict the oscillation frequencies.
We already discussed when looking at the predictions of frequency differences using the kernels that this is not always the case.
Hence, the magnitude of the deviation of the SOLA inversions due to nonlinear dependencies on the structure changes needs to be investigated.
In this case, we can compare the results of the inversions of target models in our model grid (the sum in Eq.~\ref{eqn:sola-approx}) with the actual difference in structure (the integral in Eq.~\ref{eqn:sola-approx}).
If the linear approximation is correct, then these two values should be equal.
We can therefore assess the validity of the linear approximation by plotting the inversion results and the averaged target-structure differences against each other for different target models and comparing to the diagonal $y=x$.
An example of such a comparison can be found in Fig.~\ref{fig:sola-compare-models}, where it is clear that the nonlinear dependencies cause large deviations.
Given a certain result of the inversion (the y-axis in Fig.~\ref{fig:sola-compare-models}), it is almost impossible to get a constraint on the actual structure of the star (the x-axis).
The only exception to this are negative values of the inferred difference, but this is only the case for this specific averaging kernel.
While not shown in this plot, any effect of the uncertainties on the scale of the star or the observations is much smaller than the effect of this nonlinear dependency.
For the other locations in the star and for other combinations of the oscillation frequencies, the effect is similar in size.
Hence, given that the model grid is representative of typical forward modeling uncertainties, linear inversions offer no improvement.
Looking at target model \#3, we find that while the nonlinear dependency shown in Fig.~\ref{fig:model-estimation} is quite strong, it still produced the correct result.
This may be indicative of the fact that the location of the buoyancy peak of target model \#3 is close to the reference model, compared to the peak of target model \#2 (see Fig.~\ref{fig:models-buoyancy}).
As was the case for RLS, we explore in Appendix~\ref{sec:change-number-freqs} the possibility that removing modes close to an avoided crossing may improve the quality of the inversion.
We find that this is not the case for SOLA either.
While the horizontal spread of the points in Fig.~\ref{fig:sola-compare-models} is reduced slightly, the quality of the averaging kernels is reduced considerably.

\begin{figure}
  \resizebox{\hsize}{!}{\includegraphics{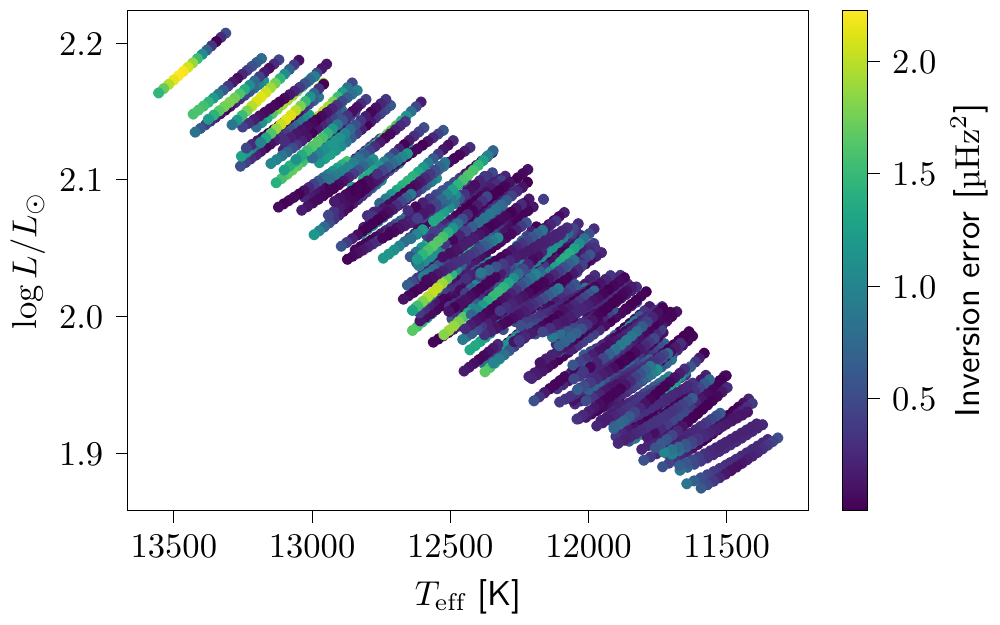}}
  \caption{Location of the evolutionary models in an HRD, as in Fig.~\ref{fig:model-sample}. Each colored circle is one of the 2880 stellar models, forming 192 evolutionary tracks. The colors indicate the error on the inversion. No significant correlation can be seen between the errors on the inferred difference in structure and the position of the model in an HRD. We note that the tracks overlap in this figure, hence the color at location of the target models is not necessarily representative of the reference models themselves.}
  \label{fig:model-sample-error}
\end{figure}

\begin{figure}
  \resizebox{\hsize}{!}{\includegraphics{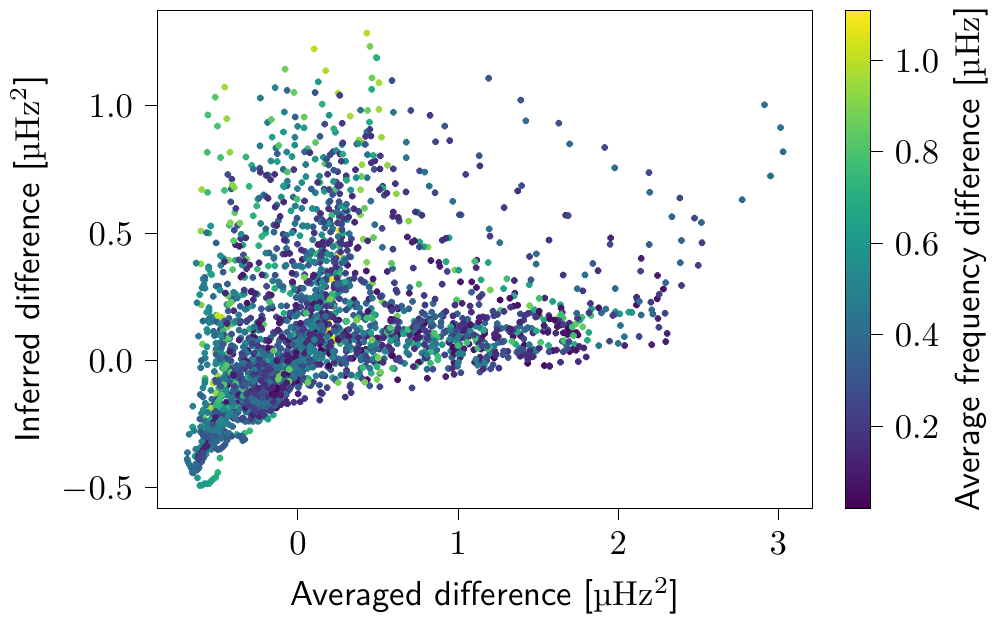}}
  \caption{Comparison of the actual difference in structure to the inferred difference as in Fig.~\ref{fig:sola-compare-models}. The colors indicate the average of the absolute frequency differences between the reference model and the target models.}
  \label{fig:sola-compare-colored-delta}
\end{figure}

These results show that current forward modeling inaccuracies are too large for accurate linear inversions.
One remaining question is then how accurate the forward modeling must be for these linear inversions to hold, or whether some observable can be used to determine whether the forward model is close enough to reality for linear inversions.
We find that there is no strong relation between the inversion error and the different model parameters (see Table~\ref{table:model-grid}), indicating that the validity region for the linear inversions cannot be represented by a simple 6D box.
This mirrors what is typically found in the case of forward modeling, where 1-sigma uncertainty regions have a highly correlated structure \citep[e.g.,][]{papics-kic-2014, michielsen-probing-2021}.
For the potential of using observables to determine whether we can infer the structure of a star using linear inversions, we considered the location of the model in an HRD (see Fig.~\ref{fig:model-sample-error}) and the average of the absolute difference between the frequencies of the reference and the target model (see Fig.~\ref{fig:sola-compare-colored-delta}).
In both cases, the inversion error is uncorrelated with the observable, excluding these observables as a way to determine whether linear inversions are valid for that particular star, given a reference model from forward modeling.

\subsection{Dependence on the number of detected modes}

A final element we considered is how much the inversion results depend on which and how many observed modes are used.
The details for this analysis can be found in Appendix~\ref{sec:change-number-freqs}.
Here we summarize the main conclusions.
We find that, as expected, reducing the number of modes included in the inversions reduces the quality of the inversions, with 20-30 modes being a reasonable amount ot still achieve meaningfull results.
Interestingly, most of the information on the internal structure of the star is found in the lower-order modes ($n= 1 -20$).
Higher-order modes do not seem to contribute substantial additional information on the properties of the near-core region.
However, it is important to take into account that these higher order modes are more affected by nonlinear effects, which were removed before testing the quality of the inversion.
Hence, correlation structures between the different oscillation modes with regards to changes in the structure of the star, could change when moving from linear inversions to a full nonlinear inversion method.

\section{Summary and conclusions}
\label{sec:conclusion}

We find that the frequencies of high-order g modes in main-sequence pulsators contain enough information about the near-core region to make inversions feasible.
Assuming that changes in the structure of the star are linearly related to changes in the oscillation frequencies allows both RLS and SOLA to accurately probe the near-core region.
While g-mode inversions require at least 20 observed and identified oscillation frequencies, preferably low-order modes, it will still be possible to perform this kind of inversion for a significant part of the known SPB stars \citep{pedersen-internal-2021}.
Furthermore, some of these stars also show quadrupole g modes ($l=2$). These have so far not yet been exploited by forward modeling and could provide more information on the internal structure, in addition to the dipole g modes.

In their current form, both RLS and SOLA do not improve on forward modeling, as the nonlinear dependency of the oscillation frequencies on the structure of the star is too strong to be handled well by the existing inversion methods. 
The main difficulty lies in the modes that are located in the dips of the period-spacing patterns of these g-mode pulsators.
In the case of our models, these dips are due to a resonance between the oscillation modes and the peak in the buoyancy frequency in the near-core region.
As this peak changes over time due to chemical evolution of the star \citep{pedersen-shape-2018,michielsen-probing-2019}, the oscillation modes will also undergo significant changes, making linear inversions challenging.
We find that removing the oscillation modes that are involved in avoided crossings does not improve the quality of the inversions appreciably.
In order to be able to invert the structure of g-mode pulsators, these nonlinear dependencies on the structure need to be included in a new inversion method dedicated to g modes in stars with a convective core.

As a next logical step, we will develop such new inversion methods.
There are multiple ways this could be accomplished: an analytical description of the avoided crossings could be developed, similar to how this has been done for g and p mixed modes \citep{ong-mixed-2021}.
Another pathway is to parameterize the buoyancy frequency profile from forward modeling, and invert for these parameters using a Markov chain Monte Carlo approach or a similar numerical approach, solving the full oscillation equations at each iteration \citep{lesaux-nonlinear-preparation}.
Since this would not use the variational principle, it would include all nonlinear dependencies on the structure in a natural way.
Finally, analytical approaches could also be used to understand what kind of buoyancy frequency profile matches with observed period-spacing patterns \citep{cunha-analytical-2019}.

Other improvements to existing inversion methods should also be considered.
Many of the known g-mode pulsators are fast rotators, causing changes to the eigenfunctions of the oscillations modes \citep{henneco-effect-2021}.
Magnetic fields may also change the oscillation frequencies, even for the zonal modes \citep{vanbeeck-detecting-2020, dhouib-detecting-2022}, which can also affect the inversion result.
Including the effects due to both rapid rotation and internal magnetism is necessary for the majority of SPB stars in order to be able to accurately determine their stellar structure from an inversion.

While the current inversion methods turn out not to be applicable to invert g modes in main-sequence pulsators in the considered mass and age regime, we were able to show that enough information is available in the oscillation frequencies to probe the near-core region.
Hence, this work motivates the future development of new inversion techniques that take these nonlinear effects into account, with the aim to unlock more information about the near-core region of g-mode main-sequence pulsators than currently available from forward modeling \citep{pedersen-diversity-2022}.

\begin{acknowledgements}
The research leading to these results has received funding from the KU Leuven Research Council (grant C16/18/005: PARADISE). V.V. and C.A. gratefully acknowledge support from the Research Foundation Flanders (FWO) under grant agreements N°1156923N (PhD Fellowship) and N°K802922N (Sabbatical leave).
Funding for the Stellar Astrophysics Centre was provided by The Danish National Research Foundation (Grant DNRF106). C.A. is grateful for the kind hospitality offered by the staff of the Center for Computational Astrophysics at the Flatiron Institute of the Simons Foundation in New York during her work visit in the fall of 2022 and spring of 2023.
This project made use of the following Python packages: NumPy \citep{harris-array-2020}, SciPy \citep{virtanen-scipy-2020}, Matplotlib \citep{hunter-matplotlib-2007}, Astropy \citep{theastropycollaboration-astropy-2013,theastropycollaboration-astropy-2018,theastropycollaboration-astropy-2022}, h5py\footnote{\url{https://www.h5py.org}}, pygyre\footnote{\url{https://pygyre.readthedocs.io}} and PyDSTool\footnote{\url{https://pydstool.github.io/PyDSTool}}.
The authors appreciated the extensive and pertinent referee report, which helped us to improve the presentation of our results. They also thank Nicholas Jannsen for proofreading and commenting on the manuscript.
\end{acknowledgements}

\bibliography{asteroseismology}

\begin{thebibliography}{104}
\expandafter\ifx\csname natexlab\endcsname\relax\def\natexlab#1{#1}\fi

\bibitem[{Aerts(2021)}]{aerts-probing-2021}
Aerts, C. 2021, Rev. Mod. Phys., 93, 015001

\bibitem[{Aerts {et~al.}(2021)Aerts, Augustson, Mathis, Pedersen, Mombarg,
  Vanlaer, Van~Beeck, \& Van~Reeth}]{aerts-rossby-2021}
Aerts, C., Augustson, K., Mathis, S., {et~al.} 2021, A\&A, 656, A121

\bibitem[{Aerts {et~al.}(2010)Aerts, {Christensen-Dalsgaard}, \&
  Kurtz}]{aerts-asteroseismology-2010}
Aerts, C., {Christensen-Dalsgaard}, J., \& Kurtz, D.~W. 2010, Asteroseismology
  ({Springer Science \& Business Media})

\bibitem[{Aerts {et~al.}(2019)Aerts, Mathis, \& Rogers}]{aerts-angular-2019}
Aerts, C., Mathis, S., \& Rogers, T.~M. 2019, Annual Review of Astronomy and
  Astrophysics, 57, 35

\bibitem[{Aerts {et~al.}(2018)Aerts, Molenberghs, Michielsen, Pedersen,
  Bj{\"o}rklund, Johnston, Mombarg, Bowman, Buysschaert, P{\'a}pics, Sekaran,
  Sundqvist, Tkachenko, Truyaert, Reeth, \& Vermeyen}]{aerts-forward-2018}
Aerts, C., Molenberghs, G., Michielsen, M., {et~al.} 2018, ApJS, 237, 15

\bibitem[{Aizenman {et~al.}(1977)Aizenman, Smeyers, \&
  Weigert}]{aizenman-avoided-1977}
Aizenman, M., Smeyers, P., \& Weigert, A. 1977, A\&A, 58, 41

\bibitem[{Aver {et~al.}(2013)Aver, Olive, Porter, \&
  Skillman}]{aver-primordial-2013}
Aver, E., Olive, K.~A., Porter, R.~L., \& Skillman, E.~D. 2013, J. Cosmol.
  Astropart. Phys., 2013, 017

\bibitem[{Basu(2003)}]{basu-stellar-2003}
Basu, S. 2003, Ap\&SS, 284, 153

\bibitem[{Basu(2016)}]{basu-global-2016}
Basu, S. 2016, Living Rev. Sol. Phys., 13, 2

\bibitem[{Basu \& Chaplin(2017)}]{basu-asteroseismic-2017}
Basu, S. \& Chaplin, W.~J. 2017, Asteroseismic Data Analysis: Foundations and
  Techniques, Vol.~4 ({Princeton University Press})

\bibitem[{Bellinger {et~al.}(2020)Bellinger, Basu, \&
  Hekker}]{bellinger-inverse-2020}
Bellinger, E.~P., Basu, S., \& Hekker, S. 2020, in Astrophysics and Space
  Science Proceedings, Vol.~57, Dynamics of the Sun and Stars; Honoring the
  Life and Work of Michael j. {{Thompson}}, ed. M.~J. P. F.~G. Monteiro, R.~A.
  Garc{\'i}a, J.~{Christensen-Dalsgaard}, \& S.~W. McIntosh, 171--183

\bibitem[{Bellinger {et~al.}(2017)Bellinger, Basu, Hekker, \&
  Ball}]{bellinger-modelindependent-2017}
Bellinger, E.~P., Basu, S., Hekker, S., \& Ball, W.~H. 2017, ApJ, 851, 80

\bibitem[{Bellinger {et~al.}(2019)Bellinger, Basu, Hekker, \&
  {Christensen-Dalsgaard}}]{bellinger-testing-2019}
Bellinger, E.~P., Basu, S., Hekker, S., \& {Christensen-Dalsgaard}, J. 2019,
  ApJ, 885, 143

\bibitem[{Bellinger {et~al.}(2021)Bellinger, Basu, Hekker,
  {Christensen-Dalsgaard}, \& Ball}]{bellinger-asteroseismic-2021}
Bellinger, E.~P., Basu, S., Hekker, S., {Christensen-Dalsgaard}, J., \& Ball,
  W.~H. 2021, ApJ, 915, 100

\bibitem[{B{\'e}trisey {et~al.}(2022)B{\'e}trisey, Pezzotti, Buldgen, Khan,
  Eggenberger, Salmon, \& Miglio}]{betrisey-kepler93-2022}
B{\'e}trisey, J., Pezzotti, C., Buldgen, G., {et~al.} 2022, A\&A, 659, A56

\bibitem[{Borucki {et~al.}(2010)Borucki, Koch, Basri, Batalha, Brown, Caldwell,
  Caldwell, {Christensen-Dalsgaard}, Cochran, DeVore, Dunham, Dupree, Thomas
  N.~Gautier, Geary, Gilliland, Gould, Howell, Jenkins, Kondo, Latham, Marcy,
  Meibom, Kjeldsen, Lissauer, Monet, Morrison, Sasselov, Tarter, Boss,
  Brownlee, Owen, Buzasi, Charbonneau, Doyle, Fortney, Ford, Holman, Seager,
  Steffen, Welsh, Rowe, Anderson, Buchhave, Ciardi, Walkowicz, Sherry, Horch,
  Isaacson, Everett, Fischer, Torres, Johnson, Endl, MacQueen, Bryson, Dotson,
  Haas, Kolodziejczak, Cleve, Chandrasekaran, Twicken, Quintana, Clarke, Allen,
  Li, Wu, Tenenbaum, Verner, Bruhweiler, Barnes, \& Prsa}]{borucki-kepler-2010}
Borucki, W.~J., Koch, D., Basri, G., {et~al.} 2010, Science

\bibitem[{Buldgen(2019)}]{buldgen-global-2019}
Buldgen, G. 2019, Bulletin de la Societe Royale des Sciences de Liege, 88, 50

\bibitem[{Buldgen {et~al.}(2022{\natexlab{a}})Buldgen, Farnir, Eggenberger,
  B{\'e}trisey, Pezzotti, Pin{\c c}on, Deal, \& Salmon}]{buldgen-thorough-2022}
Buldgen, G., Farnir, M., Eggenberger, P., {et~al.} 2022{\natexlab{a}}, A\&A,
  661, A143

\bibitem[{Buldgen {et~al.}(2022{\natexlab{b}})Buldgen, Ottoni, Pezzotti,
  Lyttle, Eggenberger, Udry, S{\'e}gransan, Miglio, Mayor, Lovis, Elsworth,
  Davies, \& Ball}]{buldgen-coralie-2022}
Buldgen, G., Ottoni, G., Pezzotti, C., {et~al.} 2022{\natexlab{b}}, A\&A, 657,
  A88

\bibitem[{Buldgen {et~al.}(2016{\natexlab{a}})Buldgen, Reese, \&
  Dupret}]{buldgen-constraints-2016}
Buldgen, G., Reese, D.~R., \& Dupret, M.~A. 2016{\natexlab{a}}, A\&A, 585, A109

\bibitem[{Buldgen {et~al.}(2017)Buldgen, Reese, \&
  Dupret}]{buldgen-analysis-2017}
Buldgen, G., Reese, D.~R., \& Dupret, M.-A. 2017, A\&A, 598, A21

\bibitem[{Buldgen {et~al.}(2019)Buldgen, Rendle, Sonoi, Davies, Miglio, Salmon,
  Reese, Bossini, Eggenberger, Noels, \& Scuflaire}]{buldgen-mean-2019}
Buldgen, G., Rendle, B., Sonoi, T., {et~al.} 2019, MNRAS, 482, 2305

\bibitem[{Buldgen {et~al.}(2016{\natexlab{b}})Buldgen, Salmon, Reese, \&
  Dupret}]{buldgen-indepth-2016}
Buldgen, G., Salmon, S. J. a.~J., Reese, D.~R., \& Dupret, M.~A.
  2016{\natexlab{b}}, A\&A, 596, A73

\bibitem[{{Christensen-Dalsgaard}(2002)}]{christensen-dalsgaard-helioseismology-2002}
{Christensen-Dalsgaard}, J. 2002, Rev. Mod. Phys., 74, 1073

\bibitem[{{Christensen-Dalsgaard}(2021)}]{christensen-dalsgaard-solar-2021}
{Christensen-Dalsgaard}, J. 2021, Living Reviews in Solar Physics, 18, 2

\bibitem[{{Christensen-Dalsgaard} {et~al.}(1990){Christensen-Dalsgaard}, Schou,
  \& Thompson}]{christensen-dalsgaard-comparison-1990}
{Christensen-Dalsgaard}, J., Schou, J., \& Thompson, M.~J. 1990, MNRAS, 242,
  353

\bibitem[{C{\'o}rsico {et~al.}(2011)C{\'o}rsico, Althaus, Kawaler,
  Miller~Bertolami, {Garc{\'i}a-Berro}, \& Kepler}]{corsico-probing-2011}
C{\'o}rsico, A.~H., Althaus, L.~G., Kawaler, S.~D., {et~al.} 2011, MNRAS, 418,
  2519

\bibitem[{Cunha {et~al.}(2019)Cunha, Avelino, {Christensen-Dalsgaard}, Stello,
  Vrard, Jiang, \& Mosser}]{cunha-analytical-2019}
Cunha, M., Avelino, P.~P., {Christensen-Dalsgaard}, J., {et~al.} 2019, MNRAS,
  490, 909

\bibitem[{Degl'Innocenti {et~al.}(2008)Degl'Innocenti, Prada~Moroni, Marconi,
  \& Ruoppo}]{deglinnocenti-franec-2008}
Degl'Innocenti, S., Prada~Moroni, P.~G., Marconi, M., \& Ruoppo, A. 2008,
  Ap\&SS, 316, 25

\bibitem[{Degroote {et~al.}(2010)Degroote, Aerts, Baglin, Miglio, Briquet,
  Noels, Niemczura, Montalban, Bloemen, Oreiro, Vu{\v c}kovi{\'c}, Smolders,
  Auvergne, Baudin, Catala, \& Michel}]{degroote-deviations-2010}
Degroote, P., Aerts, C., Baglin, A., {et~al.} 2010, Nature, 464, 259

\bibitem[{Deheuvels {et~al.}(2014)Deheuvels, Do{\u g}an, Goupil, Appourchaux,
  Benomar, Bruntt, Campante, Casagrande, Ceillier, Davies, De~Cat, Fu,
  Garc{\'i}a, Lobel, Mosser, Reese, Regulo, Schou, Stahn, Thygesen, Yang,
  Chaplin, {Christensen-Dalsgaard}, Eggenberger, Gizon, Mathis,
  {Molenda-{\.Z}akowicz}, \& Pinsonneault}]{deheuvels-seismic-2014}
Deheuvels, S., Do{\u g}an, G., Goupil, M.~J., {et~al.} 2014, A\&A, 564, A27

\bibitem[{Deheuvels {et~al.}(2012)Deheuvels, Garc{\'i}a, Chaplin, Basu, Antia,
  Appourchaux, Benomar, Davies, Elsworth, Gizon, Goupil, Reese, Regulo, Schou,
  Stahn, Casagrande, {Christensen-Dalsgaard}, Fischer, Hekker, Kjeldsen,
  Mathur, Mosser, Pinsonneault, Valenti, Christiansen, Kinemuchi, \&
  Mullally}]{deheuvels-seismic-2012}
Deheuvels, S., Garc{\'i}a, R.~A., Chaplin, W.~J., {et~al.} 2012, ApJ, 756, 19

\bibitem[{Dhouib {et~al.}(2022)Dhouib, Mathis, Bugnet, Reeth, \&
  Aerts}]{dhouib-detecting-2022}
Dhouib, H., Mathis, S., Bugnet, L., Reeth, T.~V., \& Aerts, C. 2022, A\&A, 661,
  A133

\bibitem[{Di~Mauro {et~al.}(2016)Di~Mauro, Ventura, Cardini, Stello,
  {Christensen-Dalsgaard}, Dziembowski, Patern{\`o}, Beck, Bloemen, Davies,
  Smedt, Elsworth, Garci{\textbackslash}'a, Hekker, Mosser, \&
  Tkachenko}]{dimauro-internal-2016}
Di~Mauro, M.~P., Ventura, R., Cardini, D., {et~al.} 2016, ApJ, 817, 65

\bibitem[{Di~Mauro {et~al.}(2018)Di~Mauro, Ventura, Corsaro, \&
  Moura}]{dimauro-rotational-2018}
Di~Mauro, M.~P., Ventura, R., Corsaro, E., \& Moura, B. L.~D. 2018, ApJ, 862, 9

\bibitem[{Duvall(1982)}]{duvall-dispersion-1982}
Duvall, T.~L. 1982, Nature, 300, 242

\bibitem[{Dziembowski \& Pamyatnykh(1991)}]{dziembowski-potential-1991}
Dziembowski, W.~A. \& Pamyatnykh, A.~A. 1991, A\&A, 248, L11

\bibitem[{Eggenberger {et~al.}(2008)Eggenberger, Meynet, Maeder, Hirschi,
  Charbonnel, Talon, \& Ekstr{\"o}m}]{eggenberger-geneva-2008}
Eggenberger, P., Meynet, G., Maeder, A., {et~al.} 2008, Ap\&SS, 316, 43

\bibitem[{Forestini {et~al.}(1991)Forestini, Arnould, \&
  Lumer}]{forestini-new-1991}
Forestini, M., Arnould, M., \& Lumer, E. 1991, A\&A, 252, 127

\bibitem[{Goldstein \& Townsend(2020)}]{goldstein-contour-2020}
Goldstein, J. \& Townsend, R. H.~D. 2020, ApJ, 899, 116

\bibitem[{Gough \& Thompson(1991)}]{gough-inversion-1991}
Gough, D.~O. \& Thompson, M.~J. 1991, in Solar {{Interior}} and {{Atmosphere}},
  519--561

\bibitem[{Hansen(2001)}]{hansen-lcurve-2001}
Hansen, P.~C. 2001, in Computational {{Inverse Problems}} in
  {{Electrocardiology}}, 119--142

\bibitem[{Harris {et~al.}(2020)Harris, Millman, {van der Walt}, Gommers,
  Virtanen, Cournapeau, Wieser, Taylor, Berg, Smith, Kern, Picus, Hoyer, {van
  Kerkwijk}, Brett, Haldane, {del R{\'i}o}, Wiebe, Peterson,
  {G{\'e}rard-Marchant}, Sheppard, Reddy, Weckesser, Abbasi, Gohlke, \&
  Oliphant}]{harris-array-2020}
Harris, C.~R., Millman, K.~J., {van der Walt}, S.~J., {et~al.} 2020, Nature,
  585, 357

\bibitem[{Hatta {et~al.}(2022)Hatta, Sekii, Benomar, \&
  Takata}]{hatta-bayesian-2022}
Hatta, Y., Sekii, T., Benomar, O., \& Takata, M. 2022, ApJ, 927, 40

\bibitem[{Hatta {et~al.}(2019)Hatta, Sekii, Takata, \&
  Kurtz}]{hatta-twodimensional-2019}
Hatta, Y., Sekii, T., Takata, M., \& Kurtz, D.~W. 2019, ApJ, 871, 135

\bibitem[{Henneco {et~al.}(2021)Henneco, Reeth, Prat, Mathis, Mombarg, \&
  Aerts}]{henneco-effect-2021}
Henneco, J., Reeth, T.~V., Prat, V., {et~al.} 2021, A\&A, 648, A97

\bibitem[{Hunter(2007)}]{hunter-matplotlib-2007}
Hunter, J.~D. 2007, Computing in Science \& Engineering, 9, 90

\bibitem[{Jermyn {et~al.}(2023)Jermyn, Bauer, Schwab, Farmer, Ball, Bellinger,
  Dotter, Joyce, Marchant, Mombarg, Wolf, Wong, Cinquegrana, Farrell, Smolec,
  Thoul, Cantiello, Herwig, Toloza, Bildsten, Townsend, \&
  Timmes}]{jermyn-modules-2023}
Jermyn, A.~S., Bauer, E.~B., Schwab, J., {et~al.} 2023, ApJS, 265, 15

\bibitem[{Johnston(2021)}]{johnston-one-2021}
Johnston, C. 2021, A\&A, 655, A29

\bibitem[{Kawaler {et~al.}(1999)Kawaler, Sekii, \&
  Gough}]{kawaler-prospects-1999}
Kawaler, S.~D., Sekii, T., \& Gough, D. 1999, ApJ, 516, 349

\bibitem[{Koch {et~al.}(2010)Koch, Borucki, Basri, Batalha, Brown, Caldwell,
  {Christensen-Dalsgaard}, Cochran, DeVore, Dunham, Gautier, Geary, Gilliland,
  Gould, Jenkins, Kondo, Latham, Lissauer, Marcy, Monet, Sasselov, Boss,
  Brownlee, Caldwell, Dupree, Howell, Kjeldsen, Meibom, Morrison, Owen,
  Reitsema, Tarter, Bryson, Dotson, Gazis, Haas, Kolodziejczak, Rowe,
  Van~Cleve, Allen, Chandrasekaran, Clarke, Li, Quintana, Tenenbaum, Twicken,
  \& Wu}]{koch-kepler-2010}
Koch, D.~G., Borucki, W.~J., Basri, G., {et~al.} 2010, ApJL, 713, L79

\bibitem[{Kosovichev \& Kitiashvili(2019)}]{kosovichev-resolving-2019}
Kosovichev, A.~G. \& Kitiashvili, I.~N. 2019, Proc. IAU, 15, 107

\bibitem[{Kurtz {et~al.}(2014)Kurtz, Saio, Takata, Shibahashi, Murphy, \&
  Sekii}]{kurtz-asteroseismic-2014}
Kurtz, D.~W., Saio, H., Takata, M., {et~al.} 2014, MNRAS, 444, 102

\bibitem[{Lattanzio(1986)}]{lattanzio-asymptotic-1986}
Lattanzio, J.~C. 1986, ApJ, 311, 708

\bibitem[{Le~Saux {et~al.}(2023)Le~Saux, Bellinger, \&
  Basu}]{lesaux-nonlinear-preparation}
Le~Saux, A., Bellinger, E., \& Basu, S. 2023

\bibitem[{Lee(2021)}]{lee-overstable-2021}
Lee, U. 2021, Monthly Notices of the Royal Astronomical Society, 505, 1495

\bibitem[{Lund {et~al.}(2017)Lund, Aguirre, Davies, Chaplin,
  {Christensen-Dalsgaard}, Houdek, White, Bedding, Ball, Huber, Antia,
  Lebreton, Latham, Handberg, Verma, Basu, Casagrande, Justesen, Kjeldsen, \&
  Mosumgaard}]{lund-standing-2017}
Lund, M.~N., Aguirre, V.~S., Davies, G.~R., {et~al.} 2017, ApJ, 835, 172

\bibitem[{Michielsen {et~al.}(2021)Michielsen, Aerts, \&
  Bowman}]{michielsen-probing-2021}
Michielsen, M., Aerts, C., \& Bowman, D.~M. 2021, A\&A, 650, A175

\bibitem[{Michielsen {et~al.}(2019)Michielsen, Pedersen, Augustson, Mathis, \&
  Aerts}]{michielsen-probing-2019}
Michielsen, M., Pedersen, M.~G., Augustson, K.~C., Mathis, S., \& Aerts, C.
  2019, A\&A, 628, A76

\bibitem[{Miglio {et~al.}(2008{\natexlab{a}})Miglio, Montalb{\'a}n,
  Eggenberger, \& Noels}]{miglio-gravity-2008}
Miglio, A., Montalb{\'a}n, J., Eggenberger, P., \& Noels, A.
  2008{\natexlab{a}}, Astron. Nachr., 329, 529

\bibitem[{Miglio {et~al.}(2008{\natexlab{b}})Miglio, Montalb{\'a}n, Noels, \&
  Eggenberger}]{miglio-probing-2008}
Miglio, A., Montalb{\'a}n, J., Noels, A., \& Eggenberger, P.
  2008{\natexlab{b}}, MNRAS, 386, 1487

\bibitem[{Mombarg {et~al.}(2021)Mombarg, Van~Reeth, \&
  Aerts}]{mombarg-constraining-2021}
Mombarg, J. S.~G., Van~Reeth, T., \& Aerts, C. 2021, A\&A, 650, A58

\bibitem[{Mombarg {et~al.}(2019)Mombarg, Van~Reeth, Pedersen, Molenberghs,
  Bowman, Johnston, Tkachenko, \& Aerts}]{mombarg-asteroseismic-2019}
Mombarg, J. S.~G., Van~Reeth, T., Pedersen, M.~G., {et~al.} 2019, MNRAS, 485,
  3248

\bibitem[{Moravveji {et~al.}(2015)Moravveji, Aerts, P{\'a}pics, Triana, \&
  Vandoren}]{moravveji-tight-2015}
Moravveji, E., Aerts, C., P{\'a}pics, P.~I., Triana, S.~A., \& Vandoren, B.
  2015, A\&A, 580, A27

\bibitem[{Nieva \& Przybilla(2012)}]{nieva-presentday-2012}
Nieva, M.-F. \& Przybilla, N. 2012, A\&A, 539, A143

\bibitem[{Ong \& Basu(2020)}]{ong-semianalytic-2020}
Ong, J. M.~J. \& Basu, S. 2020, ApJ, 898, 127

\bibitem[{Ong {et~al.}(2021)Ong, Basu, \& Roxburgh}]{ong-mixed-2021}
Ong, J. M.~J., Basu, S., \& Roxburgh, I.~W. 2021, ApJ, 920, 8

\bibitem[{Osaki(1975)}]{osaki-nonradial-1975}
Osaki, Y. 1975, Publications of the Astronomical Society of Japan, 27, 237

\bibitem[{Ouazzani {et~al.}(2020)Ouazzani, Ligni{\`e}res, Dupret, Salmon,
  Ballot, Christophe, \& Takata}]{ouazzani-first-2020}
Ouazzani, R.-M., Ligni{\`e}res, F., Dupret, M.-A., {et~al.} 2020, A\&A, 640,
  A49

\bibitem[{P{\'a}pics {et~al.}(2014)P{\'a}pics, Moravveji, Aerts, Tkachenko,
  Triana, Bloemen, \& Southworth}]{papics-kic-2014}
P{\'a}pics, P.~I., Moravveji, E., Aerts, C., {et~al.} 2014, A\&A, 570, A8

\bibitem[{Paxton {et~al.}(2011)Paxton, Bildsten, Dotter, Herwig, Lesaffre, \&
  Timmes}]{paxton-modules-2011}
Paxton, B., Bildsten, L., Dotter, A., {et~al.} 2011, ApJS, 192, 3

\bibitem[{Paxton {et~al.}(2013)Paxton, Cantiello, Arras, Bildsten, Brown,
  Dotter, Mankovich, Montgomery, Stello, Timmes, \&
  Townsend}]{paxton-modules-2013}
Paxton, B., Cantiello, M., Arras, P., {et~al.} 2013, ApJS, 208, 4

\bibitem[{Paxton {et~al.}(2015)Paxton, Marchant, Schwab, Bauer, Bildsten,
  Cantiello, Dessart, Farmer, Hu, Langer, Townsend, Townsley, \&
  Timmes}]{paxton-modules-2015}
Paxton, B., Marchant, P., Schwab, J., {et~al.} 2015, ApJS, 220, 15

\bibitem[{Paxton {et~al.}(2018)Paxton, Schwab, Bauer, Bildsten, Blinnikov,
  Duffell, Farmer, Goldberg, Marchant, Sorokina, Thoul, Townsend, \&
  Timmes}]{paxton-modules-2018}
Paxton, B., Schwab, J., Bauer, E.~B., {et~al.} 2018, ApJS, 234, 34

\bibitem[{Paxton {et~al.}(2019)Paxton, Smolec, Schwab, Gautschy, Bildsten,
  Cantiello, Dotter, Farmer, Goldberg, Jermyn, Kanbur, Marchant, Thoul,
  Townsend, Wolf, Zhang, \& Timmes}]{paxton-modules-2019}
Paxton, B., Smolec, R., Schwab, J., {et~al.} 2019, ApJS, 243, 10

\bibitem[{Pedersen(2022)}]{pedersen-diversity-2022}
Pedersen, M.~G. 2022, ApJ, 930, 94

\bibitem[{Pedersen {et~al.}(2021)Pedersen, Aerts, P{\'a}pics, Michielsen,
  Gebruers, Rogers, Molenberghs, Burssens, Garcia, \&
  Bowman}]{pedersen-internal-2021}
Pedersen, M.~G., Aerts, C., P{\'a}pics, P.~I., {et~al.} 2021, Nature Astronomy,
  5, 715

\bibitem[{Pedersen {et~al.}(2018)Pedersen, Aerts, P{\'a}pics, \&
  Rogers}]{pedersen-shape-2018}
Pedersen, M.~G., Aerts, C., P{\'a}pics, P.~I., \& Rogers, T.~M. 2018, A\&A,
  614, A128

\bibitem[{Pijpers \& Thompson(1994)}]{pijpers-sola-1994}
Pijpers, F.~P. \& Thompson, M.~J. 1994, A\&A, 281, 231

\bibitem[{Przybilla {et~al.}(2013)Przybilla, Nieva, Irrgang, \&
  Butler}]{przybilla-hot-2013}
Przybilla, N., Nieva, M.~F., Irrgang, A., \& Butler, K. 2013, EAS Publications
  Series, 63, 13

\bibitem[{Reese {et~al.}(2012)Reese, Marques, Goupil, Thompson, \&
  Deheuvels}]{reese-estimating-2012}
Reese, D.~R., Marques, J.~P., Goupil, M.~J., Thompson, M.~J., \& Deheuvels, S.
  2012, A\&A, 539, A63

\bibitem[{Ricker {et~al.}(2014)Ricker, Winn, Vanderspek, Latham, Bakos, Bean,
  {Berta-Thompson}, Brown, Buchhave, Butler, Butler, Chaplin, Charbonneau,
  {Christensen-Dalsgaard}, Clampin, Deming, Doty, Lee, Dressing, Dunham, Endl,
  Fressin, Ge, Henning, Holman, Howard, Ida, Jenkins, Jernigan, Johnson,
  Kaltenegger, Kawai, Kjeldsen, Laughlin, Levine, Lin, Lissauer, MacQueen,
  Marcy, McCullough, Morton, Narita, Paegert, Palle, Pepe, Pepper, Quirrenbach,
  Rinehart, Sasselov, Sato, Seager, Sozzetti, Stassun, Sullivan, Szentgyorgyi,
  Torres, Udry, \& Villasenor}]{ricker-transiting-2014}
Ricker, G.~R., Winn, J.~N., Vanderspek, R., {et~al.} 2014, JATIS, 1, 014003

\bibitem[{Roxburgh {et~al.}(1998)Roxburgh, Audard, Basu,
  {Christensen-Dalsgaard}, \& Vorontsov}]{roxburgh--1998}
Roxburgh, I.~W., Audard, N., Basu, S., {Christensen-Dalsgaard}, J., \&
  Vorontsov, S.~V. 1998, in Proc. {{IAU Symp}}. 181: {{Sounding Solar}} and
  {{Stellar Interiors}} (Poster Vol.), {Nice Observatory}, 245

\bibitem[{Saio {et~al.}(2021)Saio, Takata, Lee, Li, \&
  Van~Reeth}]{saio-rotation-2021}
Saio, H., Takata, M., Lee, U., Li, G., \& Van~Reeth, T. 2021, MNRAS, 502, 5856

\bibitem[{Scuflaire(1974)}]{scuflaire-non-1974}
Scuflaire, R. 1974, Astronomy and Astrophysics, Vol. 36, p. 107 (1974), 36, 107

\bibitem[{Seaton(2005)}]{seaton-opacity-2005}
Seaton, M.~J. 2005, Monthly Notices of the Royal Astronomical Society: Letters,
  362, L1

\bibitem[{Shibahashi {et~al.}(1988)Shibahashi, Sekii, \&
  Kawaler}]{shibahashi-white-1988}
Shibahashi, H., Sekii, T., \& Kawaler, S. 1988, in Atmospheric {{Diagnostics}}
  of {{Stellar Evolution}}: {{Chemical Peculiarity}}, {{Mass Loss}}, and
  {{Explosion}}, ed. K.~Nomoto, Lecture {{Notes}} in {{Physics}} ({Springer}),
  86--87

\bibitem[{Smeyers \& Van~Hoolst(2010)}]{smeyers-linear-2010}
Smeyers, P. \& Van~Hoolst, T. 2010, Astrophysics and {{Space Science Library}},
  Vol. 371, Linear {{Isentropic Oscillations}} of {{Stars}} ({Berlin,
  Heidelberg}: {Springer})

\bibitem[{Szewczuk {et~al.}(2022)Szewczuk, Walczak,
  {Daszy{\'n}ska-Daszkiewicz}, \& Mo{\'z}dzierski}]{szewczuk-seismic-2022}
Szewczuk, W., Walczak, P., {Daszy{\'n}ska-Daszkiewicz}, J., \& Mo{\'z}dzierski,
  D. 2022, MNRAS, 511, 1529

\bibitem[{Takata \& Montgomery(2001)}]{takata-seismic-2001}
Takata, M. \& Montgomery, M. 2001, in {{ASP Conference Series}}, Vol. 259,
  Radial and {{Nonradial Pulsations}} as {{Probes}} of {{Stellar Physics}}, ed.
  C.~Aerts, T.~R. Bedding, \& J.~{Christensen-Dalsgaard} ({Astronomical Society
  of the Pacific}), 606--607

\bibitem[{Tassoul(1980)}]{tassoul-asymptotic-1980}
Tassoul, M. 1980, ApJS, 43, 469

\bibitem[{{The Astropy Collaboration} {et~al.}(2022){The Astropy
  Collaboration}, {Price-Whelan}, Lim, Earl, Starkman, Bradley, Shupe, Patil,
  Corrales, Brasseur, N{\"o}the, Donath, Tollerud, Morris, Ginsburg, Vaher,
  Weaver, Tocknell, Jamieson, van Kerkwijk, Robitaille, Merry, Bachetti,
  G{\"u}nther, Authors, Aldcroft, {Alvarado-Montes}, Archibald, B{\'o}di,
  Bapat, Barentsen, Baz{\'a}n, Biswas, Boquien, Burke, Cara, Cara, Conroy,
  Conseil, Craig, Cross, Cruz, D'Eugenio, Dencheva, Devillepoix, Dietrich,
  Eigenbrot, Erben, Ferreira, {Foreman-Mackey}, Fox, Freij, Garg, Geda,
  Glattly, Gondhalekar, Gordon, Grant, Greenfield, Groener, Guest, Gurovich,
  Handberg, Hart, {Hatfield-Dodds}, Homeier, Hosseinzadeh, Jenness, Jones,
  Joseph, Kalmbach, Karamehmetoglu, Ka{\l}uszy{\'n}ski, Kelley, Kern,
  Kerzendorf, Koch, Kulumani, Lee, Ly, Ma, MacBride, Maljaars, Muna, Murphy,
  Norman, O'Steen, Oman, Pacifici, Pascual, {Pascual-Granado}, Patil, Perren,
  Pickering, Rastogi, Roulston, Ryan, Rykoff, Sabater, Sakurikar, Salgado,
  Sanghi, Saunders, Savchenko, Schwardt, {Seifert-Eckert}, Shih, Jain, Shukla,
  Sick, Simpson, Singanamalla, Singer, Singhal, Sinha, Sip{\H o}cz, Spitler,
  Stansby, Streicher, {\v S}umak, Swinbank, Taranu, Tewary, Tremblay,
  de~{Val-Borro}, Kooten, Vasovi{\'c}, Verma, Cardoso, Williams, Wilson,
  Winkel, {Wood-Vasey}, Xue, Yoachim, Zhang, Zonca, \&
  Contributors}]{theastropycollaboration-astropy-2022}
{The Astropy Collaboration}, {Price-Whelan}, A.~M., Lim, P.~L., {et~al.} 2022,
  ApJ, 935, 167

\bibitem[{{The Astropy Collaboration} {et~al.}(2018){The Astropy
  Collaboration}, {Price-Whelan}, Sip{\H o}cz, G{\"u}nther, Lim, Crawford,
  Conseil, Shupe, Craig, Dencheva, Ginsburg, VanderPlas, Bradley,
  {P{\'e}rez-Su{\'a}rez}, de~{Val-Borro}, Contributors), Aldcroft, Cruz,
  Robitaille, Tollerud, Committee), Ardelean, Babej, Bach, Bachetti, Bakanov,
  Bamford, Barentsen, Barmby, Baumbach, Berry, Biscani, Boquien, Bostroem,
  Bouma, Brammer, Bray, Breytenbach, Buddelmeijer, Burke, Calderone,
  Rodr{\'i}guez, Cara, Cardoso, Cheedella, Copin, Corrales, Crichton, D'Avella,
  Deil, Depagne, Dietrich, Donath, Droettboom, Earl, Erben, Fabbro, Ferreira,
  Finethy, Fox, Garrison, Gibbons, Goldstein, Gommers, Greco, Greenfield,
  Groener, Grollier, Hagen, Hirst, Homeier, Horton, Hosseinzadeh, Hu, Hunkeler,
  Ivezi{\'c}, Jain, Jenness, Kanarek, Kendrew, Kern, Kerzendorf, Khvalko, King,
  Kirkby, Kulkarni, Kumar, Lee, Lenz, Littlefair, Ma, Macleod, Mastropietro,
  McCully, Montagnac, Morris, Mueller, Mumford, Muna, Murphy, Nelson, Nguyen,
  Ninan, N{\"o}the, Ogaz, Oh, Parejko, Parley, Pascual, Patil, Patil, Plunkett,
  Prochaska, Rastogi, Janga, Sabater, Sakurikar, Seifert, Sherbert,
  {Sherwood-Taylor}, Shih, Sick, Silbiger, Singanamalla, Singer, Sladen,
  Sooley, Sornarajah, Streicher, Teuben, Thomas, Tremblay, Turner, Terr{\'o}n,
  van Kerkwijk, de~la Vega, Watkins, Weaver, Whitmore, Woillez, Zabalza, \&
  Contributors)}]{theastropycollaboration-astropy-2018}
{The Astropy Collaboration}, {Price-Whelan}, A.~M., Sip{\H o}cz, B.~M.,
  {et~al.} 2018, AJ, 156, 123

\bibitem[{{The Astropy Collaboration} {et~al.}(2013){The Astropy
  Collaboration}, Robitaille, Tollerud, Greenfield, Droettboom, Bray, Aldcroft,
  Davis, Ginsburg, {Price-Whelan}, Kerzendorf, Conley, Crighton, Barbary, Muna,
  Ferguson, Grollier, Parikh, Nair, G{\"u}nther, Deil, Woillez, Conseil,
  Kramer, Turner, Singer, Fox, Weaver, Zabalza, Edwards, Bostroem, Burke,
  Casey, Crawford, Dencheva, Ely, Jenness, Labrie, Lim, Pierfederici, Pontzen,
  Ptak, Refsdal, Servillat, \&
  Streicher}]{theastropycollaboration-astropy-2013}
{The Astropy Collaboration}, Robitaille, T.~P., Tollerud, E.~J., {et~al.} 2013,
  A\&A, 558, A33

\bibitem[{Tokuno \& Takata(2022)}]{tokuno-asteroseismology-2022}
Tokuno, T. \& Takata, M. 2022, MNRAS, 514, 4140

\bibitem[{Townsend {et~al.}(2018)Townsend, Goldstein, \&
  Zweibel}]{townsend-angular-2018}
Townsend, R. H.~D., Goldstein, J., \& Zweibel, E. 2018, MNRAS, 475, 879

\bibitem[{Townsend \& Teitler(2013)}]{townsend-gyre-2013}
Townsend, R. H.~D. \& Teitler, S.~A. 2013, MNRAS, 435, 3406

\bibitem[{Triana {et~al.}(2017)Triana, Corsaro, Ridder, Bonanno, Hern{\'a}ndez,
  \& Garc{\'i}a}]{triana-internal-2017}
Triana, S.~A., Corsaro, E., Ridder, J.~D., {et~al.} 2017, A\&A, 602, A62

\bibitem[{Triana {et~al.}(2015)Triana, Moravveji, P{\'a}pics, Aerts, Kawaler,
  \& {Christensen-Dalsgaard}}]{triana-internal-2015}
Triana, S.~A., Moravveji, E., P{\'a}pics, P.~I., {et~al.} 2015, ApJ, 810, 16

\bibitem[{Van~Beeck {et~al.}(2021)Van~Beeck, Bowman, Pedersen, Van~Reeth,
  Van~Hoolst, \& Aerts}]{vanbeeck-detection-2021}
Van~Beeck, J., Bowman, D.~M., Pedersen, M.~G., {et~al.} 2021, A\&A, 655, A59

\bibitem[{Van~Beeck {et~al.}(2020)Van~Beeck, Prat, Reeth, Mathis, Bowman,
  Neiner, \& Aerts}]{vanbeeck-detecting-2020}
Van~Beeck, J., Prat, V., Reeth, T.~V., {et~al.} 2020, A\&A, 638, A149

\bibitem[{Van~Reeth {et~al.}(2015)Van~Reeth, Tkachenko, Aerts, P{\'a}pics,
  Triana, Zwintz, Degroote, Debosscher, Bloemen, Schmid, De~Smedt, Fremat,
  Fuentes, Homan, Hrudkova, Karjalainen, Lombaert, Nemeth, {\O}stensen, Van
  De~Steene, Vos, Raskin, \& Van~Winckel}]{vanreeth-gravitymode-2015}
Van~Reeth, T., Tkachenko, A., Aerts, C., {et~al.} 2015, ApJS, 218, 27

\bibitem[{Virtanen {et~al.}(2020)Virtanen, Gommers, Oliphant, Haberland, Reddy,
  Cournapeau, Burovski, Peterson, Weckesser, Bright, {van der Walt}, Brett,
  Wilson, Millman, Mayorov, Nelson, Jones, Kern, Larson, Carey, Polat, Feng,
  Moore, VanderPlas, Laxalde, Perktold, Cimrman, Henriksen, Quintero, Harris,
  Archibald, Ribeiro, Pedregosa, \& {van Mulbregt}}]{virtanen-scipy-2020}
Virtanen, P., Gommers, R., Oliphant, T.~E., {et~al.} 2020, Nat Methods, 17, 261

\bibitem[{Weiss \& Schlattl(2008)}]{weiss-garstec-2008}
Weiss, A. \& Schlattl, H. 2008, Ap\&SS, 316, 99

\end{thebibliography}

\begin{appendix}
    \section{Kernels for \texorpdfstring{$(N^2, \rho)$ and $(N^2, c)$}{(N2, rho) and (N2, c)}}
    \label{sec:kernel-derivations}

    The kernels for the variable pairs $(\rho, c)$ and $(N^2, \rho)$ are related by
    \begin{equation} \int_0^{R}\left[ K_{c,\rho}\frac{\delta c}{c} + K_{\rho,c}\frac{\delta \rho}{\rho}\right]\dd{r} = \int_0^{R}\left[ K_{\rho,N^2}\frac{\delta \rho}{\rho} + K_{N^2,\rho}\delta N^2\right]\dd{r}\,. \end{equation}
    As the perturbation in terms of $\delta \rho$ is independent from $\delta c$, we should find that the factors in terms of $\delta \rho$ and $\delta c$ are the same on both sides.
    We therefore expand the perturbation of the buoyancy frequency to first order in the density and the sound speed perturbations:
    \begin{equation}
    \begin{aligned}
            \delta N^2 &= \delta \left[g\left(\frac{1}{\Gamma_1} \dv{\ln p}{r} - \dv{ \ln \rho}{r}\right)\right] \\
                               &= \frac{\delta g}{g} N^2 + \frac{g}{c^2 \rho} \dv{\delta p }{r} - 2 \frac{g}{c^2 \rho}\dv{p}{r} \frac{\delta c}{c}\\
                               &\quad- \frac{g}{c^2 \rho}\dv{p}{r} \frac{\delta \rho}{\rho} + g\dv{\ln \rho}{r} \frac{\delta\rho}{\rho} - g\frac{1}{\rho} \dv{\delta\rho}{r}\,.
    \end{aligned}
    \label{eqn:inversion:delta-N2}\end{equation}
    We further expand this to $\delta \rho$ and $\delta c$ using partial integration and by dropping all surface terms.
    This leads to the following relation between the kernels:
    \begin{equation}
    \begin{aligned}
            K_{c,\rho} &= 2\frac{g^2}{c^2}K_{N^2,\rho}\,, \\
            K_{\rho, c} &= K_{N^2,\rho}\dv{g}{r}
            + 4 \pi G r^2 \rho \int_r^{R} K_{N^2,\rho} s^{-2}\left(\frac{N^2}{g} - \frac{g}{c^2}\right)\dd{s} \\
            &+ g\dv{}{r}K_{N^2,\rho} + K_{\rho,N^2}\,.
    \end{aligned}\label{eqn:kernel-relation-rho}
    \end{equation}
    From the first equation we obtain $K_{N^2,\rho}$, which we then substitute in the second equation to obtain $K_{\rho,N^2}$. 

    The equations for the $(N^2, c)$ kernels are derived in a similar way, only one term is moved:
    \begin{equation}
    \begin{aligned}
            K_{c,\rho} &= 2\frac{g^2}{c^2}K_{N^2,c} + K_{c,N^2}\,, \\
            K_{\rho, c} &= K_{N^2,c}\dv{g}{r}
            + 4 \pi G r^2 \rho \int_r^{R} K_{N^2,c} s^{-2}\left(\frac{N^2}{g} - \frac{g}{c^2}\right)\dd{s} \\
            &+ g\dv{}{r}K_{N^2,c}\,.
    \end{aligned}\label{eqn:kernel-relation-c}
    \end{equation}
    These equations cannot be solved easily.
    The second equation is a integro-differential equation for $K_{N^2,c}$ and needs to be integrated numerically.
    This can be accomplished by introducing a new variable $U$ equaling the integral in Eq.~\eqref{eqn:kernel-relation-c} and rewriting for $\dv{}{r}U$ and $\dv{}{r}K_{N^2, c}$:
    \begin{equation}
      \begin{aligned}
        &\dv{}{r}U = -K_{N^2, c} r^{-2} \left(\frac{N^2}{g} - \frac{g}{c^2}\right) \\
        &\dv{}{r}K_{N^2, c} = g^{-1}K_{\rho, c} +  \dv{\log g}{r}\,K_{N^2,c} + r^2\dv{\log m}{r}\,U\,.
      \end{aligned}
    \end{equation}
    This equation can be integrated, given some boundary conditions for $U$ and $K_{N^2,c}$.
    At the surface we have the condition that $U = 0$ as the upper and lower limit of the integral are then the same.
    We choose $K_{N^2,c} = 0$ as well at the surface, although the choice of this boundary condition should not affect the computed frequency difference from these kernels, but may cause the contributions to shift between the buoyancy frequency and the sound speed.
    Varying the boundary condition to be nonzero shows, however, that in all cases the zero boundary conditions suppress the sound speed component maximally.

\begin{figure}
       \resizebox{\hsize}{!}{\includegraphics{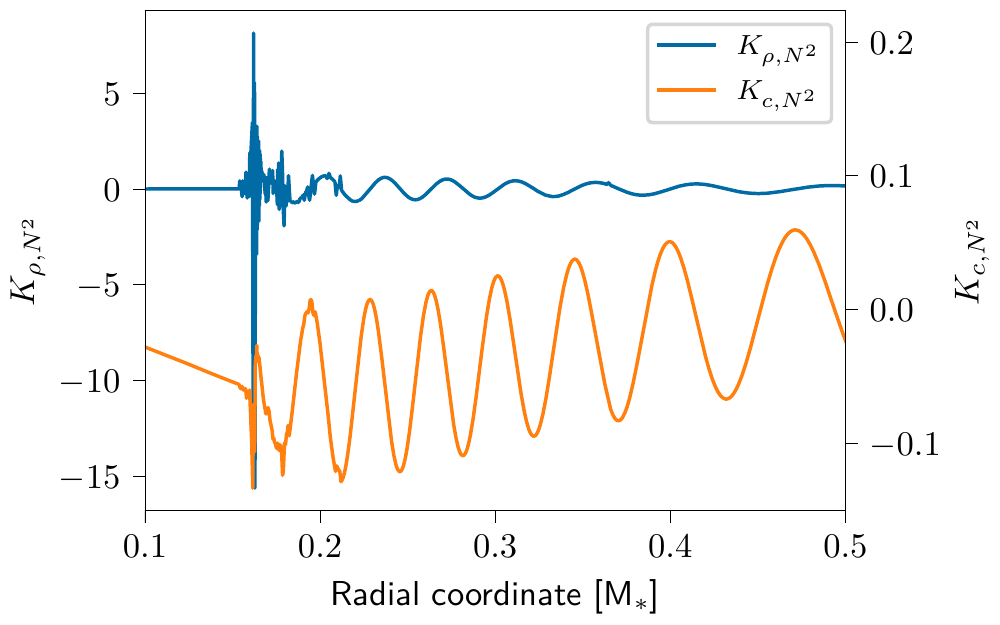}}
       \caption{Kernels for the $n = 10,l = 1$ g mode of the reference model. Both kernels do show some numerical glitches, but they are significantly worse for the $K_{\rho,N^2}$ kernel. The amplitude of the $K_{\rho,N^2}$ is also two orders of magnitude large than the $K_{c,N^2}$ kernel.}
       \label{fig:noisy-kernels}
    \end{figure}

    As was shown in Sect.~\ref{sec:test}, the kernel pair $(N^2,\rho)$ is worse at estimating the oscillation frequencies of the target models.
    This could be caused by numerical issues arising in the computation of these kernels.
    We explore this by plotting the $K_{\rho,N^2}$ and $K_{c,N^2}$ kernels in Fig.~\ref{fig:noisy-kernels}.
    Here we find that the $K_{c,N^2}$ kernel is much better behaved than the $K_{\rho,N^2}$.
    While the $K_{c,N^2}$ kernel also contributes almost nothing to the overall predicted frequency difference, the $K_{\rho,N^2}$ kernel does not, propagating these numerical errors to the predicted differences.
    We suspect that the $K_{c,N^2}$ kernel is better behaved because the numerical integration scheme used to solve Eq.~\eqref{eqn:kernel-relation-c} is more stable against numerical errors than the differentiation scheme used to solve Eq.~\eqref{eqn:kernel-relation-rho}.
    As we found that the $(N^2,c)$ variable pair is an excellent choice for an inversion, we did not dedicate further time to resolving these numerical errors.

    \section{Reducing the effect of the scaling parameter}
    \label{sec:reduced-scaling}

\begin{figure}
    \resizebox{\hsize}{!}{\includegraphics{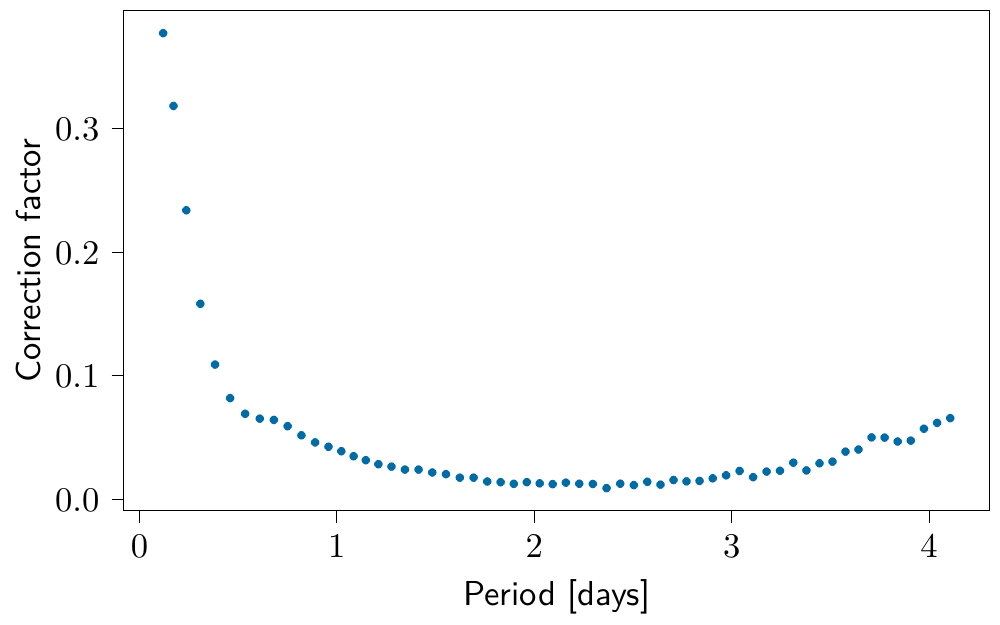}}
    \caption{Correction factor to the term $\delta q / q$, given by $1 - \int_0^R 2N^2 q^2 K_{N^2, c}\dd{r}$, for the reference model. This shows how the effect of the scaling is reduced by transforming from dimensionless $N^2$ to dimensional $N^2$.}
    \label{fig:scaling-reduced}
  \end{figure}

    The effect of the scaling parameter on the predictions and the inversion can be reduced by transforming $\delta N^2$ from dimensionless form to a dimensional form.
    We start from Eq.~\eqref{eqn:kernel-definition} for the variable pair $(N^2, c)$, ignoring the $c$ term:
    \begin{equation} \frac{\delta \omega}{\omega} + \frac{\delta q}{q} = \int_0^R \delta \mathcal{N}^2 K_{N^2, c}\dd{r}\end{equation}
    where we use $\mathcal{N}$ to refer to the dimensionless variant of the buoyancy frequency.
    This can then be rewritten using the relation $N^2 q^2 = \mathcal{N}^2$ since the buoyancy frequency has dimensions of [time$^{-2}$]:
    \begin{equation}
      \begin{aligned}
        \frac{\delta \omega}{\omega} + \frac{\delta q}{q} &= \int_0^R \delta (N^2 q^2) K_{N^2, c}\dd{r}  \\
                                                          &= \int_0^R \left((\delta N^2) q^2 + N^2 (\delta q^2)\right) K_{N^2, c}\dd{r}
      \end{aligned}
    \end{equation}
    where we drop the term $(\delta N^2)(\delta q^2)$.
    Transforming $\delta q^2 = 2q\delta q$ and moving the $\delta q$ terms together, we find that
    \begin{equation} \frac{\delta \omega}{\omega} + \left(1 - \int_0^R 2N^2 q^2 K_{N^2, c}\dd{r} \right) \frac{\delta q}{q} = \int_0^R \delta N^2 q^2 K_{N^2, c}\dd{r}\,. \end{equation}
    As long as the integral over $2N^2 q^2 K_{N^2, c}$ is close to one, the effect of the scaling parameter will be greatly reduced.
    Figure~\ref{fig:scaling-reduced} shows that this is indeed the case, especially for oscillation modes with a high radial order.

  \section{Changing the number of detected frequencies}
  \label{sec:change-number-freqs}
  The goal of this section is to investigate how the inversion methods behave when less oscillation modes are used.

  \subsection{Shorter series of modes}

\begin{figure*}
     \resizebox{\hsize}{!}{\includegraphics{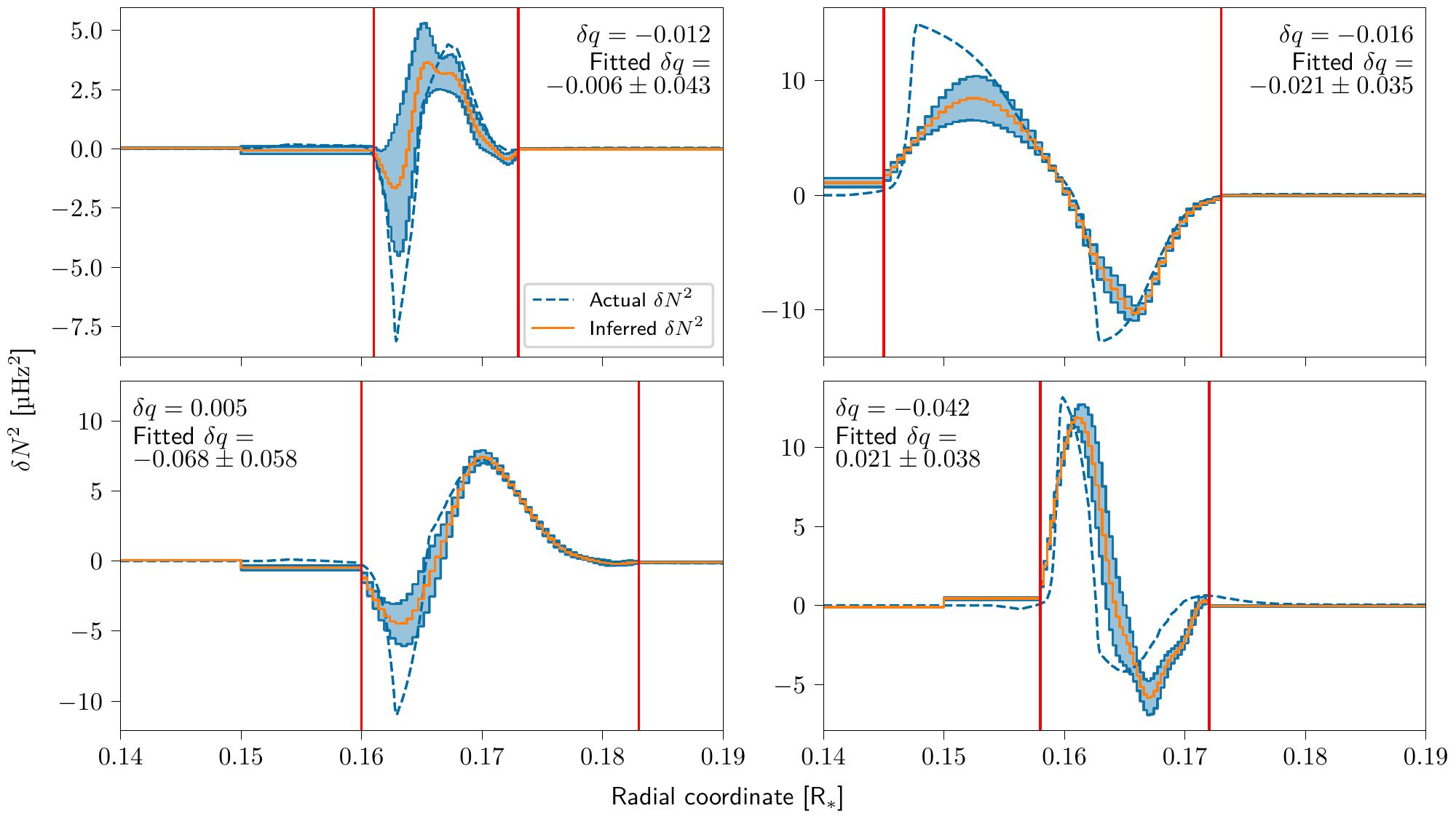}}
     \caption{Structure inversions of the four target models using RLS, relying on oscillation modes with $n = 10-50$. The inferred buoyancy frequency difference is shown in orange (full line) and the difference between the models in blue (dashed line). The observational uncertainty is indicated by the light blue regions around the inferred profile. The inferred profile is discretized in 70 grid points, of which 50 are located in regions of the largest difference in buoyancy frequency, indicated by the red lines. The first row contains target models \#1 and \#2, while the second row contains target models \#3 and \#4. As in Fig.~\ref{fig:rls-4-targets-corrected}, the difference in oscillation frequencies between the target models is computed from the difference in structure and the kernels.}
     \label{fig:rls-40-freqs-predicted}
  \end{figure*}

\begin{figure*}
     \resizebox{\hsize}{!}{\includegraphics{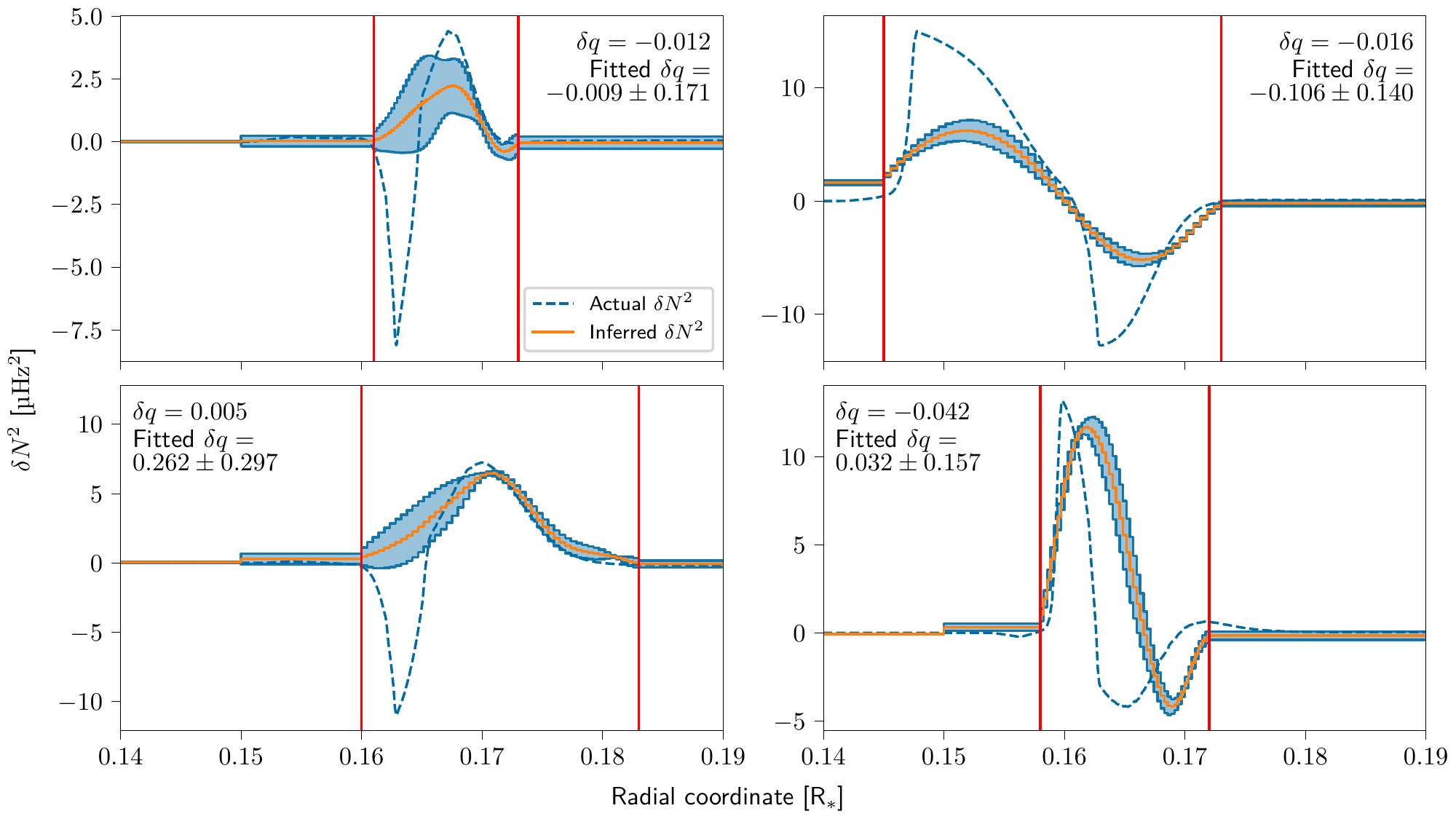}}
     \caption{Same as Fig.~\ref{fig:rls-40-freqs-predicted}, but with oscillation modes of radial order $n=20-40$.}
     \label{fig:rls-20-freqs-predicted}
  \end{figure*}

  In Sect.~\ref{sec:test}, only the case of all dipole frequencies from $n = 1$ to $n = 60$ being used in the inversions was considered, with the aim to focus on the nonlinear structure dependencies.
  However, typical g-mode pulsators tend to have (significantly) less clearly identified modes.
  Hence, it is important to know whether information about the internal structure of the star can be recovered with less modes.
  We test the properties of the inversion methods with the following set of modes: $n = 10-50$ and $20-40$ for RLS and SOLA, and additionally $n= 1-20$ and $10-30$ for SOLA to show the impact of the lower-order modes.
  Since we are not considering the effects of the avoided crossings here, we use frequency differences obtained from the structure differences and the kernels, as for Fig.~\ref{fig:rls-4-targets-corrected}.
  The results for RLS are shown in Fig.~\ref{fig:rls-40-freqs-predicted} and Fig.~\ref{fig:rls-20-freqs-predicted}.
  The inversion parameters are the same as in Fig.~\ref{fig:rls-4-targets-corrected}, except for the regularization parameter, which has been decreased to account for the reduction in the number of oscillation modes.
  We find that the inferred profiles are similar to the actual differences, as was the case with the full set of modes.
  However, some details are missing as is expected when less modes are considered.
  Overall, the observational uncertainties are larger, although this depends on the exact value of the regularization parameter.
  The larger the regularization parameter, typically the less the uncertainties on the frequencies are amplified by the inversion.
  The change in the scaling parameter $q$ is significantly different for target models \#2-4 and the uncertainty is much larger.
  This could be caused by the reduced sensitivity of the higher-order modes to the scaling parameter, as was shown in Appendix~\ref{sec:reduced-scaling}.

\begin{figure}
     \resizebox{\hsize}{!}{\includegraphics{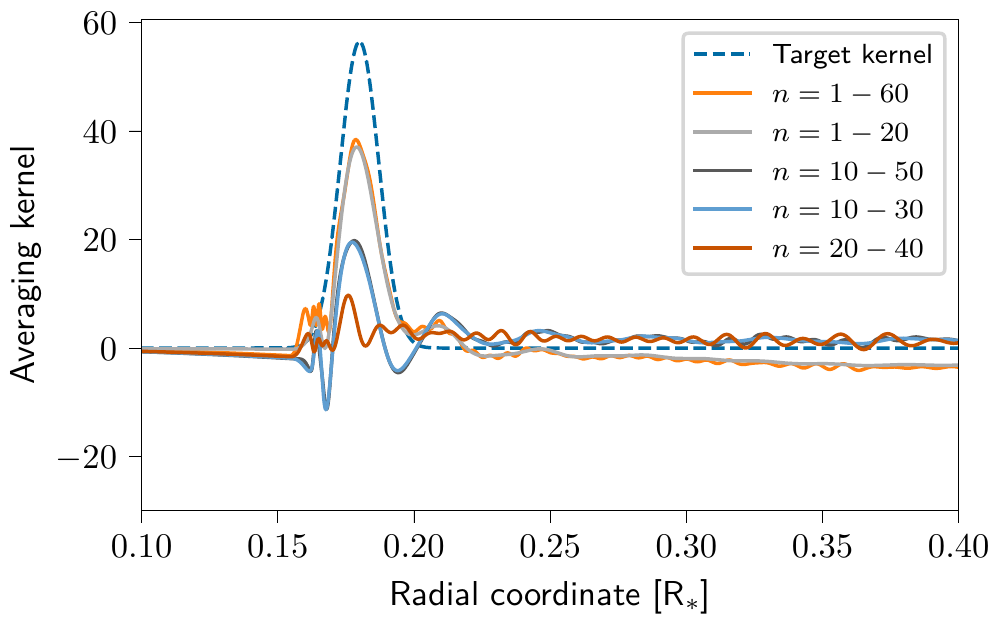}}
     \caption{Target and averaging kernel for the reference model localized just above the near-core region, for multiple sets of oscillation modes. The target kernel used is a modified Gaussian located at $r = 0.18$ and a width of 0.01. The inversion parameters have been tuned to obtain an uncertainty of $\sim$\SI{0.1}{\micro\hertz\squared} as in Figs.~\ref{fig:sola-near-core} and \ref{fig:sola-outside-near-core}.}
     \label{fig:sola-limit-modes}
  \end{figure}

\begin{figure}
     \resizebox{\hsize}{!}{\includegraphics{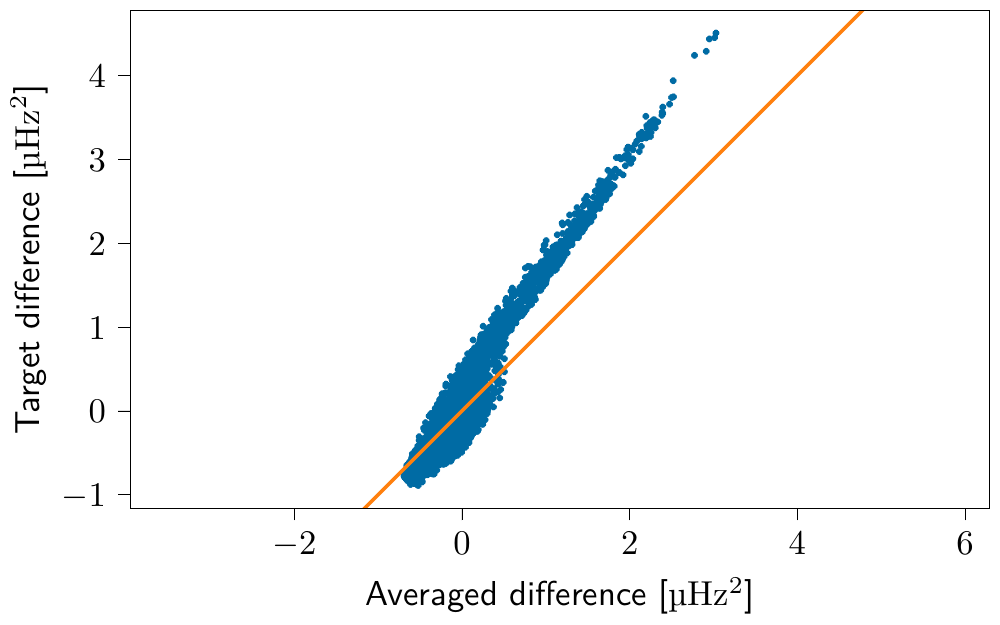}}
     \caption{Quality of the averaging kernel for the $n=1-60$ averaging kernel from Fig.~\ref{fig:sola-limit-modes}. The x axis shows the difference between the reference model and the 2880 target models averaged by the averaging kernel, while the y axis shows the same difference averaged by the target kernel. The closer the points lie to the $y=x$ line (in orange), the better the averaging kernel matches the target kernel. The numbers indicate the different target models.}
     \label{fig:sola-limit-modes-compare-1}
  \end{figure}

\begin{figure}
     \resizebox{\hsize}{!}{\includegraphics{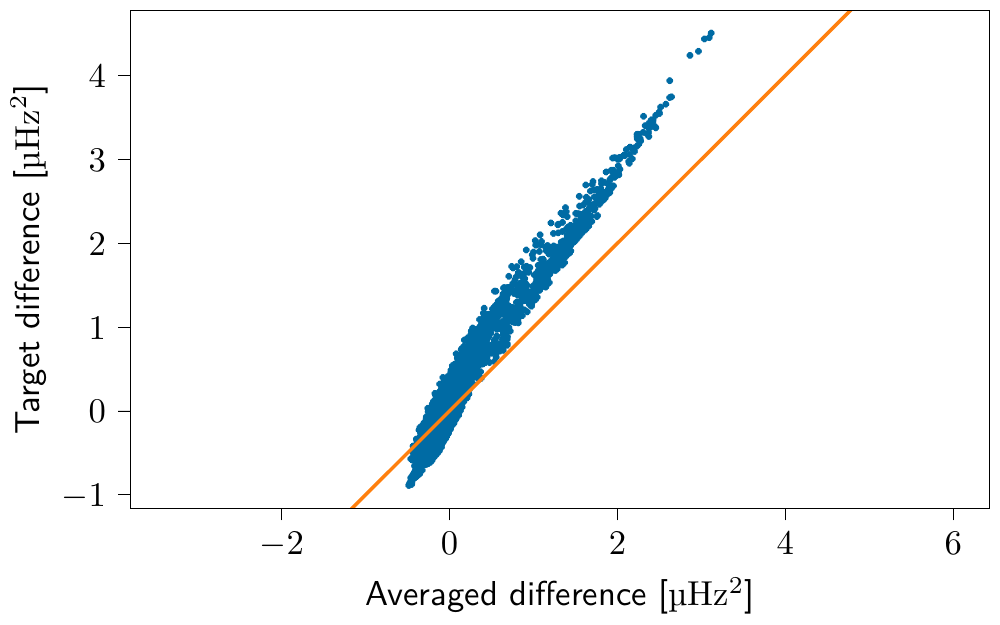}}
     \caption{Same as Fig.~\ref{fig:sola-limit-modes-compare-1}, but with the $n=1-20$ mode set.}
     \label{fig:sola-limit-modes-compare-2}
  \end{figure}

\begin{figure}
     \resizebox{\hsize}{!}{\includegraphics{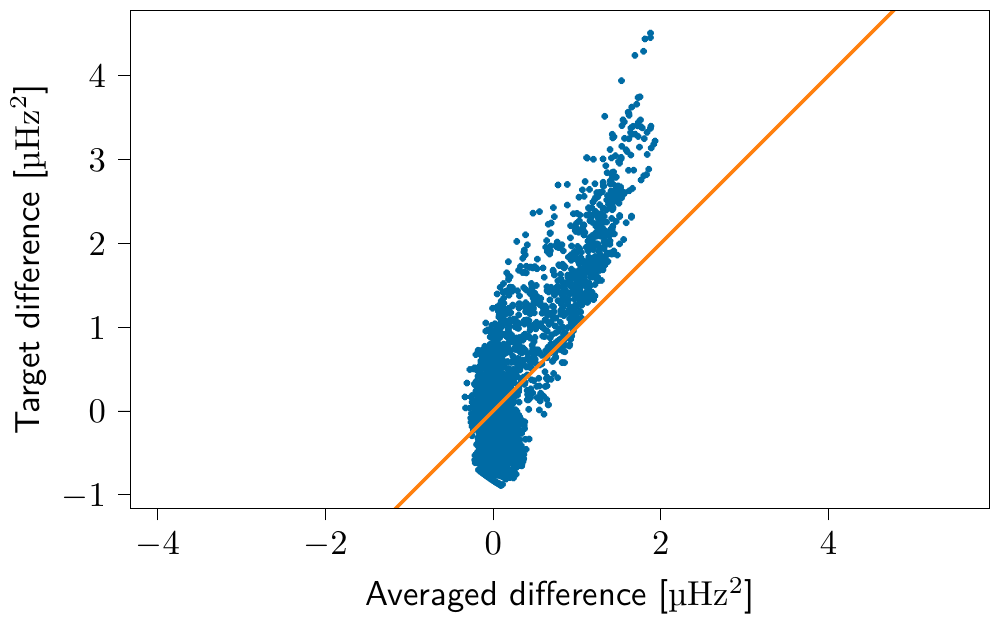}}
     \caption{Same as Fig.~\ref{fig:sola-limit-modes-compare-1}, but with the $n=10-50$ mode set.}
     \label{fig:sola-limit-modes-compare-3}
  \end{figure}

\begin{figure}
     \resizebox{\hsize}{!}{\includegraphics{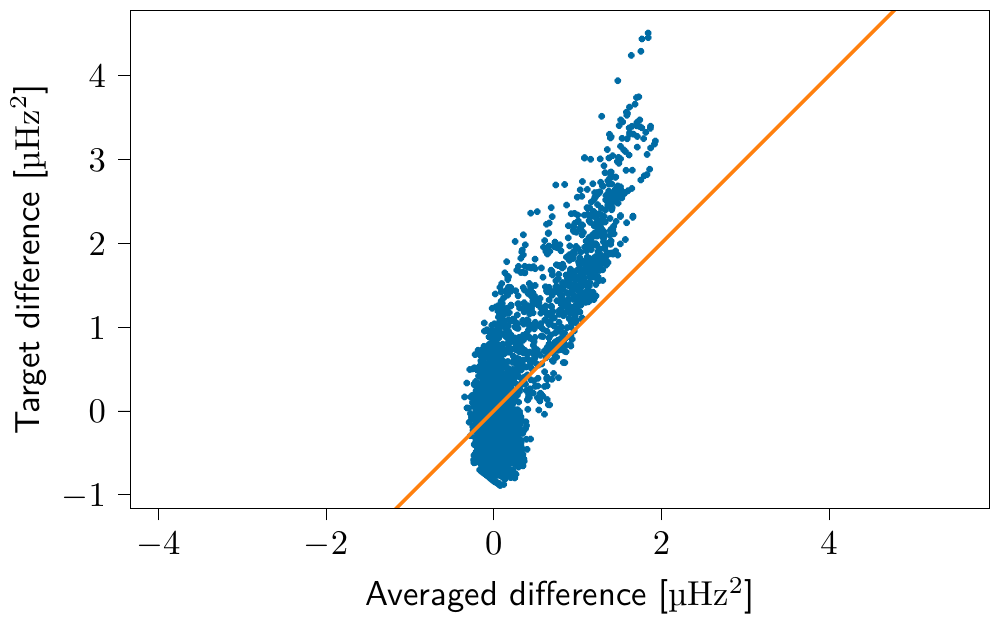}}
     \caption{Same as Fig.~\ref{fig:sola-limit-modes-compare-1}, but with the $n=10-30$ mode set.}
     \label{fig:sola-limit-modes-compare-4}
  \end{figure}

\begin{figure}
     \resizebox{\hsize}{!}{\includegraphics{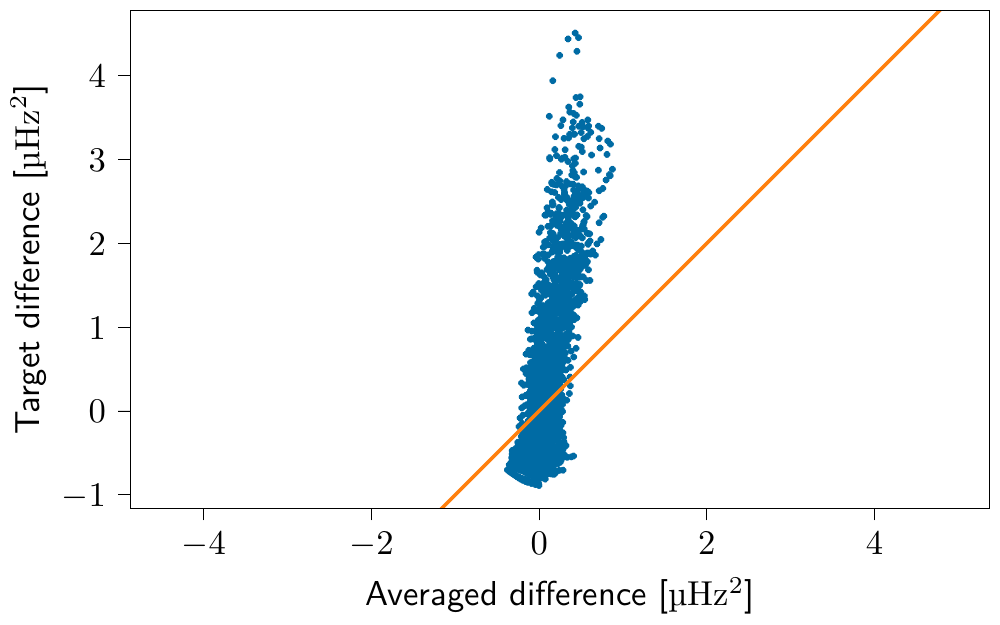}}
     \caption{Same as Fig.~\ref{fig:sola-limit-modes-compare-1}, but with the $n=20-40$ mode set.}
     \label{fig:sola-limit-modes-compare-5}
  \end{figure}

  For SOLA, we plot the averaging kernels for the different sets of models in Fig.~\ref{fig:sola-limit-modes}, computed using the same procedure as in Sect.~\ref{sec:test}.
  Once more, we are not considering the effect of the avoided crossings, just the information content in the oscillation modes.
  We use the same modified Gaussian target kernel centered at $r = 0.18$ and with a width of $0.01$, as was used in Fig.~\ref{fig:sola-outside-near-core}.
  For this target kernel, the exclusion of the low-order modes ($n = 1-10$) seems to have the largest impact on the quality of the averaging kernels.
  Comparing the cases $n = 10-30$ and $n = 10-50$, we find that including more higher-order modes even slightly deteriorates the quality of the averaging kernel, introducing more local oscillations.
  The overall shape of the averaging kernel is barely impacted, however.
  This indicates either that the additional modes do not contain extra information about that region in the star that is not already contained in the other modes, or that SOLA and/or the target kernel are not optimal for this use case.
  The quality of these kernels can also be visualized using the model grid.
  In Figs.~\ref{fig:sola-limit-modes-compare-1} to \ref{fig:sola-limit-modes-compare-5} we compare for each of the 2880 target models the difference in structure averaged by the averaging kernel and averaged by the target kernel.
  For the full set of modes ($n=1-60$), the difference from the averaging kernel is close to the difference from the target kernel, indicating that the averaging kernel is well localized.
  We do note that there is still some difference in the slope between this full set of modes and the diagonal, as the averaging kernel still has some sidelobes outside of the region of the target kernel.
  As most of the changes to the structure are in the near-core region, these sidelobes contribute little to the averaged difference.
  Therefore, the averaged difference is typically be lower than the target difference.
  In line with the averaging kernels being similar, using only modes $n=1-20$ leads to similar results as can be seen in Fig.~\ref{fig:sola-limit-modes-compare-2}.
  When the first 10 lower-order modes are dropped, this relation is less clean, see Figs.~\ref{fig:sola-limit-modes-compare-3} and \ref{fig:sola-limit-modes-compare-4}.
  If 10 more low-order modes are removed, the averaging kernel essentially has no peak anymore at the location of the target kernel, and the relation between the two is almost completely gone (see Fig.~\ref{fig:sola-limit-modes-compare-5}).
  Therefore, for this target kernel, the most beneficial modes are the lower-order modes ($n=1-20$), while the number of higher-order modes is not that relevant.

  \subsection{Removing modes around the avoided crossings}

\begin{figure*}
     \resizebox{\hsize}{!}{\includegraphics{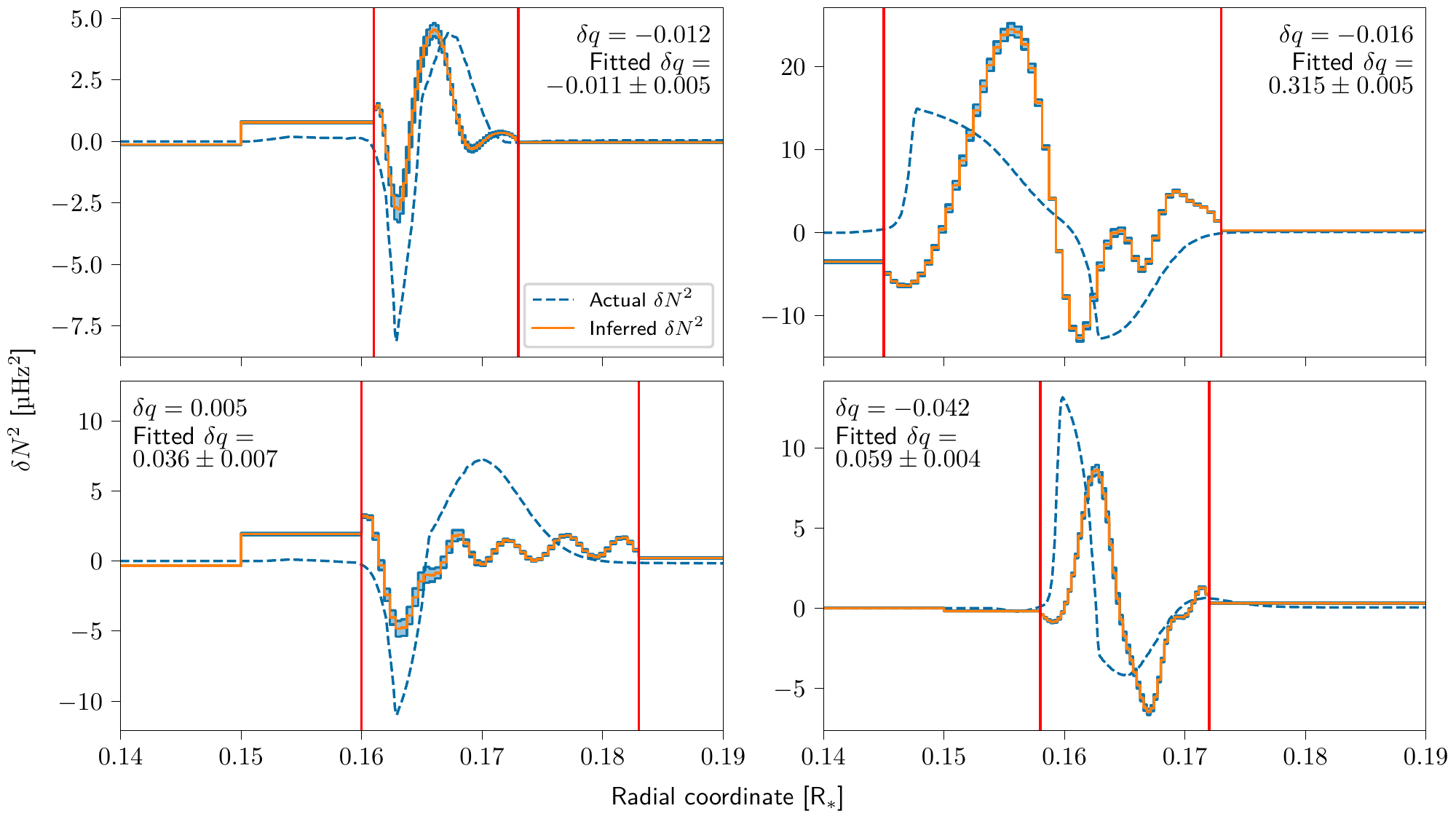}}
     \caption{Same as Fig.~\ref{fig:rls-40-freqs-predicted}, but with oscillation modes around the avoided crossings removed, leaving 40 modes for the inversions and using the frequencies as computed by GYRE.}
     \label{fig:rls-cutout-freqs-predicted-3}
  \end{figure*}

\begin{figure*}
     \resizebox{\hsize}{!}{\includegraphics{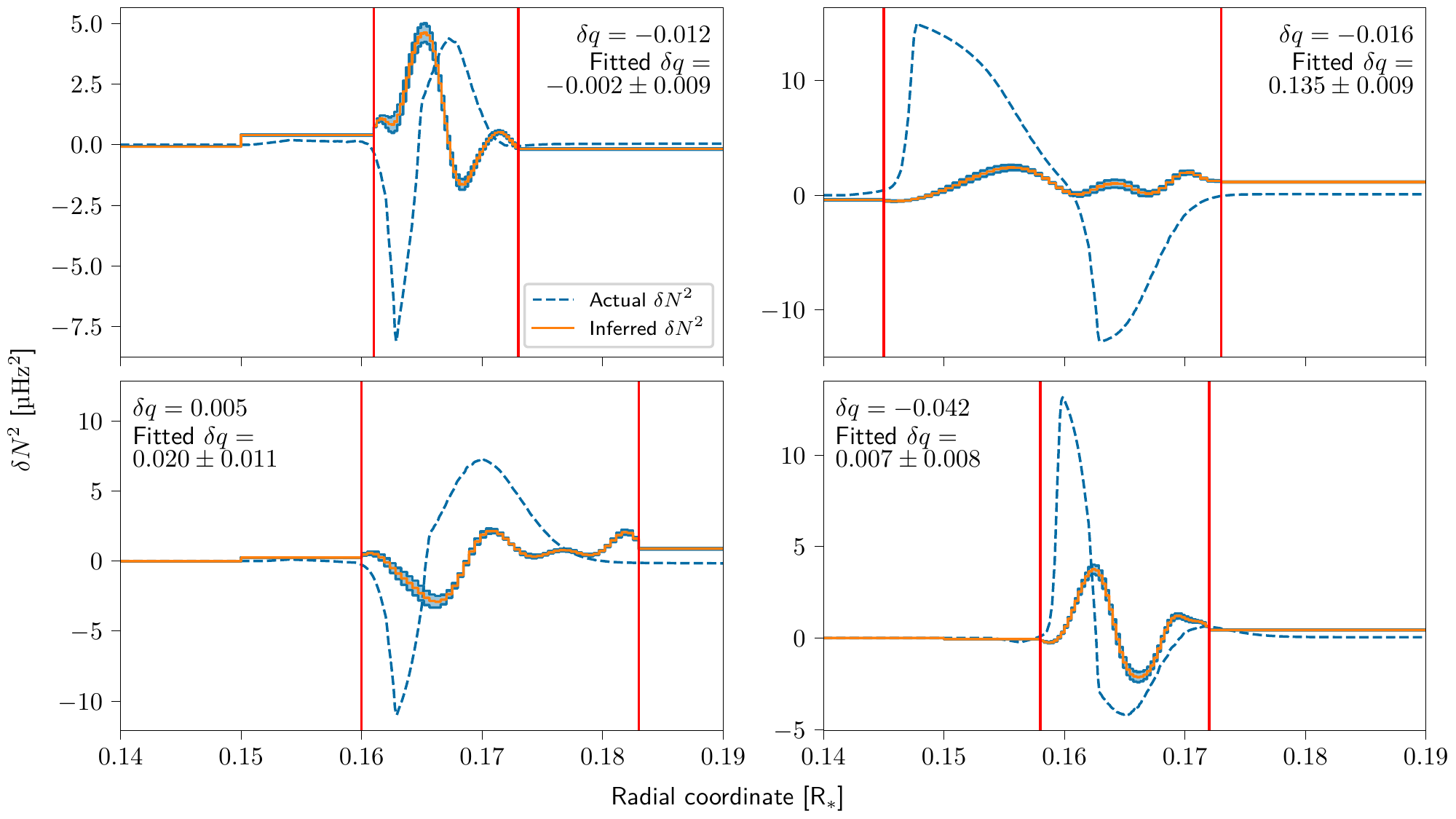}}
     \caption{Same as Fig.~\ref{fig:rls-40-freqs-predicted}, but with oscillation modes around the avoided crossings removed, leaving 24 modes for the inversions and using the frequencies as computed by GYRE.}
     \label{fig:rls-cutout-freqs-predicted-5}
  \end{figure*}

\begin{figure}
     \resizebox{\hsize}{!}{\includegraphics{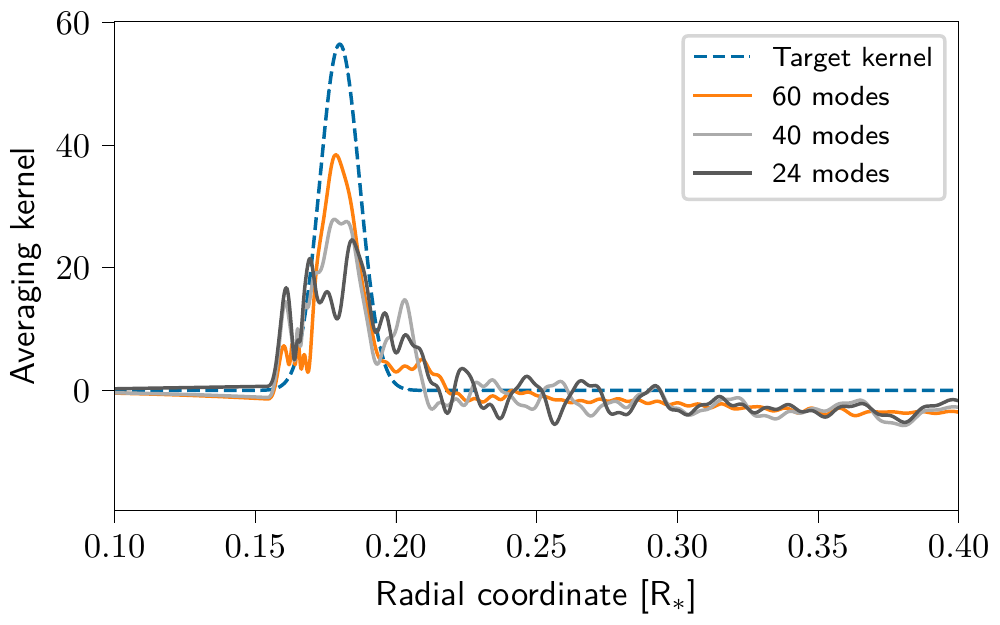}}
     \caption{Same as Fig.~\ref{fig:sola-limit-modes}, but in the cases with 40 and 24 modes we have removed respectively two or four modes on each side of the avoided crossings.}
     \label{fig:sola-cutout-kernels}
  \end{figure}

\begin{figure}
     \resizebox{\hsize}{!}{\includegraphics{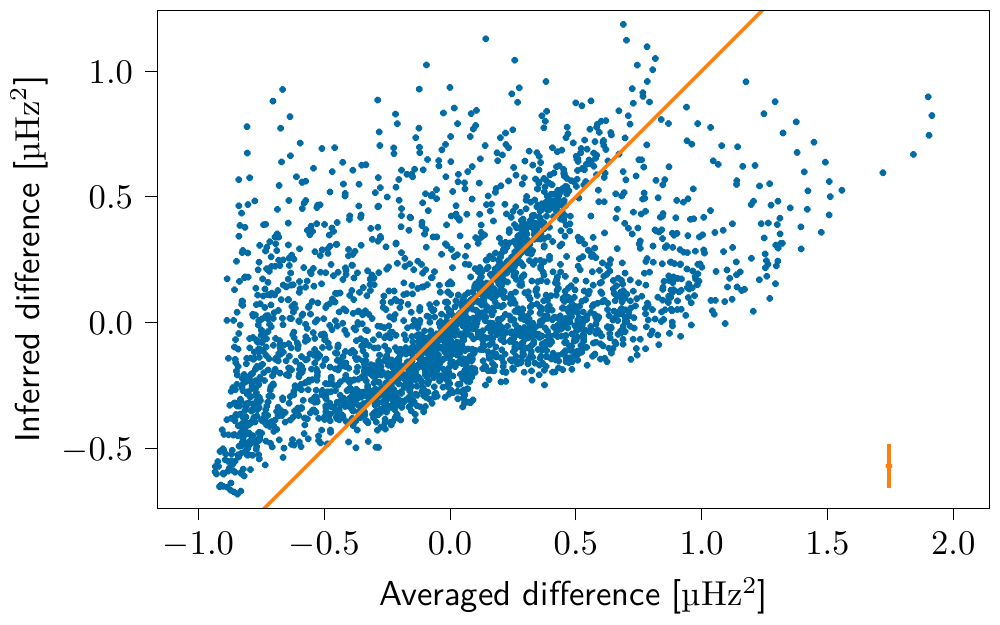}}
    \caption{Comparison of the actual difference in structure to the inferred difference for the 40 mode averaging kernel in Fig.~\ref{fig:sola-cutout-kernels}. The values on the $x$- and $y$-axis are differences in squared buoyancy frequency. The $x$-axis shows the integral of the structure difference multiplied by the averaging kernel (the integral in Eq.~\ref{eqn:sola-approx}), while the $y$-axis shows the inversion result (the left hand side of Eq.~\ref{eqn:sola-approx}). Each blue dot is one of the 2880 stellar models. The orange diagonal line is given by $x = y$. All deviations from this line are caused by the nonlinear dependency on the structure. The numbers indicate the different target models. The orange error bar in the lower right corner of the plot indicates the observational uncertainty on the inferred difference, given an uncertainty of $10^{-3}$\SI{}{\micro\hertz} on the observed frequencies.}
     \label{fig:sola-cutout-compare-40}
  \end{figure}

\begin{figure}
     \resizebox{\hsize}{!}{\includegraphics{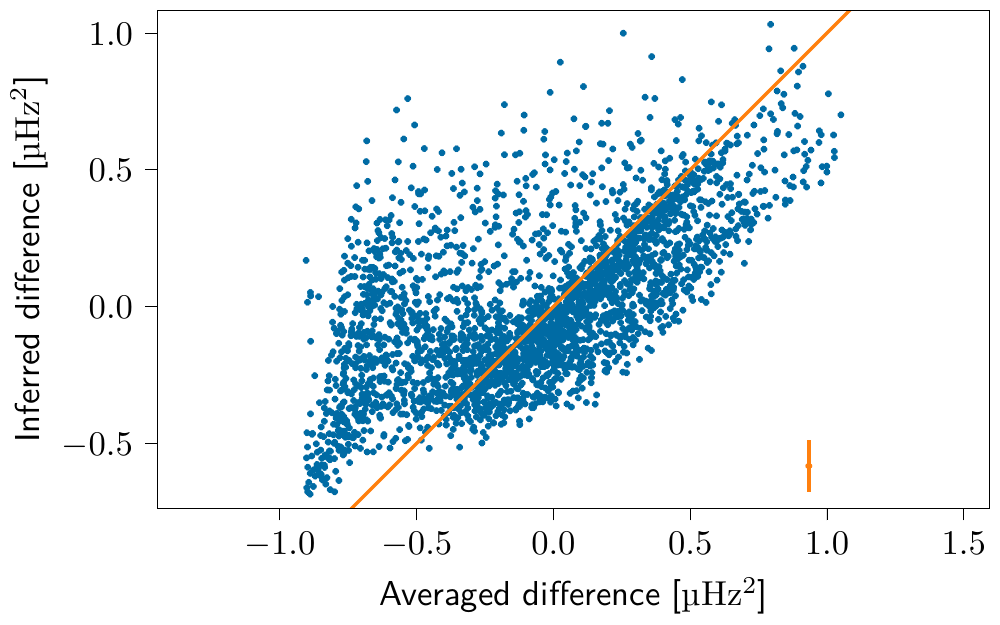}}
     \caption{Same as Fig.~\ref{fig:sola-cutout-compare-40}, but with the 24 mode averaging kernel instead.}
     \label{fig:sola-cutout-compare-24}
  \end{figure}

  Another goal of this section is to test whether removing the modes in avoided crossings at the location of the reference model can reduce the nonlinear dependencies on the structure.
  However, as these modes couple with the peak in the buoyancy frequency, they might provide a significant amount of information about that peak.
  Hence, for this to work, it is necessary to balance the loss of information and the reduction of the nonlinear dependency.
  We identify at which radial order the avoided crossings take place based on Fig.~\ref{fig:avoided-crossings} and remove those modes along with a certain number of modes around these selected modes.
  For the reference model, we find that the avoided crossings take place at $n=10, 23, 34$ and $45$.
  We remove these modes and two or four modes at each side of these specific modes, that is we remove modes with radial orders $n= 8-12,21-25,32-36,43-47$ or $n= 6-14,19-27,30-38,41-49$, leaving 40 and 24 modes respectively.
  Contrary to the previous section, we do not use the modified frequencies as predicted by the kernels, but the actual frequencies as computed by GYRE.
  The results of the inversions with RLS can be found in Fig.~\ref{fig:rls-cutout-freqs-predicted-3} and Fig.~\ref{fig:rls-cutout-freqs-predicted-5}.
  As in the previous section, the regularization parameter has been decreased.
  For the first set of modes, the results appear qualitatively similar, but certain features are now absent.
  For example, the dip in the difference next to the core for target models \#1 and \#3 is less pronounced compared to Fig.~\ref{fig:rls-4-targets}.
  However, the mismatches due to the nonlinear dependency in the structure in target models \#2 to 4 are not significantly reduced.
  In the case of the second set of modes, the inferred difference no longer resembles the actual difference, indicating that too much information is lost by removing those frequencies.
  However, the locations of the additional peaks and dips in the profiles caused by the nonlinear dependencies from Fig.~\ref{fig:rls-cutout-freqs-predicted-3} can still be seen in Fig.~\ref{fig:rls-cutout-freqs-predicted-5} for models \#2 to 4.
  These features are smaller, but still present.
  The uncertainties on these results are substantially smaller than in the previous section (Figs.~\ref{fig:rls-40-freqs-predicted} and \ref{fig:rls-20-freqs-predicted}), where similar numbers of modes were selected.
  The results of the same sets of modes, but now using SOLA inversions can be found in Fig.~\ref{fig:sola-cutout-kernels}, showing the averaging kernel and comparisons between the inferred difference and the averaged difference.
  We attribute this to the absence of the low order modes in the $n= 10-50$ and $20-60$ mode sets, as the tests with SOLA showed that these carry the most information about the near-core region.
  The width and location of the averaging kernel has been kept the same as for Fig.~\ref{fig:sola-outside-near-core}.
  As expected, the localization of the averaging kernel is reduced by the removal of the modes.
  Figures~\ref{fig:sola-cutout-compare-40} and \ref{fig:sola-cutout-compare-24} compare the inferred difference in the buoyancy frequency with the actual difference averaged by the averaging kernel for the two sets of modes.
  Similar to the case of RLS, the nonlinear features, namely the spread of the points away from the $x=y$ line, are reduced, but still present.
  Given an inferred difference, it is still difficult to determine what the real difference should be.
  We therefore conclude that removing modes around the avoided crossings does not increase the quality of the inferred profiles.
  However, the nonlinearities can still be caused by the avoided crossings, as we have only removed modes around the avoided crossings of the reference model.
  As the properties of the star change, so too will the avoided crossings move in radial order or frequency.
  Consider the case of the core hydrogen fraction as was shown in Fig.~\ref{fig:avoided-crossings}.
  Just the range $0.6 > X_c > 0.575$ alone has almost all modes in avoided crossings at some point, except for some of the lower order modes.

\end{appendix}
\end{document}